\documentclass[useAMS,usenatbib]{mn2e}
\usepackage{subfigure}
\usepackage{graphicx}
\usepackage{amssymb}
\usepackage{pdflscape}
\usepackage[normalem]{ulem}

\title[The IRXCS of type 1 AGN]{The infrared to X-ray correlation spectra of unobscured type 1 active galactic nuclei}
\author[I. Garc\'ia-Bernete et al.]
{\parbox{\textwidth}{I. Garc\'ia-Bernete,$^{1,2}$\thanks{E-mail: igarcia@iac.es} C. Ramos Almeida,$^{1,2}$ H. Landt,$^{3}$ M. J. Ward,$^{3}$ M. Balokovi\'c$^{4}$ and J. A. Acosta-Pulido$^{1,2}$.}\vspace{0.4cm}\\
\parbox{\textwidth}{$^{1}$Instituto de Astrof\'isica de Canarias, Calle V\'ia L\'actea, s/n, E-38205, La Laguna, Tenerife, Spain\\
$^{2}$Departamento de Astrof\'isica, Universidad de La Laguna, E-38206, La Laguna, Tenerife, Spain\\
$^{3}$Centre for Extragalactic Astronomy, Department of Physics, Durham University, South Road, Durham DH1 3LE, UK\\
$^{4}$Cahill Center for Astronomy and Astrophysics, California Institute of Technology, 1216 E California Blvd, Pasadena, CA 91125, USA\\
}
}
\begin{document}
\date{}
\pagerange{\pageref{firstpage}--\pageref{lastpage}} \pubyear{2017}
\maketitle
\label{firstpage}
\begin{abstract}
We use new X-ray data obtained with the {\textit{Nuclear Spectroscopic Telescope Array}} ({\textit{NuSTAR}}), near-infrared (NIR) fluxes, and mid-infrared (MIR) spectra of a sample of 24 unobscured type 1 active galactic nuclei (AGN) to study the correlation between various hard X-ray bands between 3 and 80~keV and the infrared (IR) emission. The IR to X-ray correlation spectrum (IRXCS) shows a maximum at $\sim$15-20~$\mu$m, coincident with the peak of the AGN contribution to the MIR spectra of the majority of the sample. There is also a NIR correlation peak at $\sim$2~$\mu$m, which we associate with the NIR bump observed in some type 1 AGN at $\sim$1--5 $\mu$m and is likely produced by nuclear hot dust emission. The IRXCS shows practically the same behaviour in all the X-ray bands considered, indicating a common origin for all of them. We finally evaluated correlations between the X-ray luminosities and various MIR emission lines. All the lines show a good correlation with the hard X-rays ($\rho\ge$0.7), but we do not find the expected correlation between their ionization potentials and the strength of the IRXCS. 
\end{abstract}


\section{Introduction}
\label{intro}
Active galactic nuclei (AGN) are powered by accretion of material onto supermassive black holes (SMBHs), which release energy in the form of radiation and/or mechanical outflows into the host galaxy's interstellar medium. This feedback process appears to be fundamental to the formation and evolution of galaxies \citep{Hopkins10}. Therefore, it is important to characterize the properties of AGN in the local universe to understand how they are triggered and whether all galaxies with SMBHs go through an active phase. 

Due to the high energies involved in the accretion process, AGN are strong X-ray emitters. This emission is mainly produced by the Comptonization of accretion disc photons in a hot corona of electrons surrounding the SMBH (e.g., \citealt{Haardt91}). On the other hand, the unified model of AGN proposes that there is dust surrounding the active nucleus distributed in a toroidal geometry \citep{Antonucci1993} which obscures the central engines of type 2 AGN, and allows a direct view in the case of type 1 sources. Previous X-ray studies confirmed this scheme since, in general, type 2s have higher absorption column densities than type 1 AGN (e.g., \citealt{Awaki91,Smith96,Turner97,Bassani99,Cappi06,Dadina08,Ricci11,Singh11}). Although some exceptions have been observed (e.g., \citealt{Cappi06,deRosa07,Corral11}), those are expected if the broad line region (BLR) and the dusty torus have a clumpy distribution (see e.g., \citealt{Elitzur12} and references therein). 

High energy X-ray observations of active galaxies enable studies of the intrinsic emission from the central engine since: 1) they are less sensitive to the effects of obscuration than softer X-ray energies, and 2) very high energies are involved in the accretion process. The main source of X-ray emission is the intrinsic AGN continuum, which is observed from $\sim$1~keV to over 100~keV. This primary X-ray emission can be reflected (e.g., inverse Compton 
scattering of photons from the accretion disc; \citealt{Jovanovic08}) and the main features of this reflection component are the so-called ``Compton hump'', which peaks at $\sim$30~keV \citep{George91}, and the Fe K$_{\alpha}$ fluorescence line at 6.4~keV. The Compton hump is produced by the reprocessing of X-ray photons by Compton-thick material, but the exact location of such material (the corona, the BLR and/or the torus) is not clear.

The dusty torus absorbs the intrinsic AGN radiation, and then reprocesses it to emerge in the infrared (IR), peaking in the mid-IR (MIR; $\sim$5-30~$\mu$m) according to torus models (e.g., \citealt{Pier92}). Thus, MIR observations of active galaxies are key to study the emission of dust heated by the AGN, but also by star-formation (SF) when present (e.g., \citealt{Radomski2003,Packham05,Sales13,Esquej14,Herrero14,Ramos14,Ruschel14,Bernete15}). The main spectral features of AGN in the MIR are the silicates, the polycyclic aromatic hydrocarbon (PAH) emission bands, and several emission lines of different ionization potential (IP). 

The PAH features are often used to measure the star formation rate (SFR) of galaxies (see e.g., \citealt{Peeters04,Wu05,Diamond12,Esquej14}), together with low IP MIR emission lines such as [Ne\,II]$\lambda$12.81~$\mu$m and [S\,III]$\lambda$18.71~$\mu$m \citep{Spinoglio92,Spinoglio12,Ho07,Pereira-Santaella2010b}. On the other hand, the presence and strengths of high IP emission lines such as [Ne\,V]$\lambda$14.32~$\mu$m ($\sim$97~eV) and [O\,IV]$\lambda$25.89~$\mu$m ($\sim$55~eV) are considered to be reliable indicators of the AGN power. The [O\,IV] emission line has proved to be a reliable AGN tracer (see e.g., \citealt{Bernete16} and references therein), which correlates well with both the hard X-rays \citep{Melendez08,Rigby09,Diamond09} and the soft X-rays \citep{Prieto02}. Another AGN tracer commonly used is the [S\,IV]$\lambda$10.51~$\mu$m line (IP$\sim$35~eV; \citealt{Dasyra11}). However, this emission line can also be produced in star forming regions, as shown by \citet{Pereira-Santaella2010a}. The same applies to the [Ne\,III]$\lambda$15.56~$\mu$m (IP$\sim$41~eV) emission line \citep{Ho07,Gorjian07,Melendez08,Pereira-Santaella2010b}.

In view of the apparent connection between the AGN's high energy continuum and some IR features, it is of interest to examine this in detail. Although the MIR--X-ray correlation has been extensively studied in the literature \citep{Krabbe01, Prieto02, Lutz04, Ramos07, Horst08, Fiore09, Gandhi09, Levenson09, Mason12, Sazonov12, Matsuta12, Ichikawa12, Asmus2015, Mateos15, Bernete16, Ichikawa16, Chen17}, to date there have been no detailed studies using the entire NIR-to-MIR range and selected X-ray bands from 3 to 80~keV. The aim of this work is to investigate this correlation for a sample of 24 nearby unobscured type 1 AGN using new X-ray data obtained with the {\textit{Nuclear Spectroscopic Telescope Array}} ({\textit{NuSTAR}}; \citealt{Harrison13}) together with archival IR data.

The paper is organized as follows. Section \ref{sample} and Section \ref{observations} describe the sample selection and the observations, respectively. The X-ray and MIR spectral modelling are presented in Section \ref{modelling}. Section \ref{correlation_spectrum} describes the correlation spectrum technique. The main results on the IR to X-ray correlations are presented in Section \ref{results}. Finally, in Section \ref{Discussion} we present the discussion and in Section \ref{Conclusions} we summarize the main conclusions of this work. 

Throughout this paper we assumed a cosmology with H$_0$=73 km~s$^{-1}$~Mpc$^{-1}$, $\Omega_m$=0.27, and $\Omega_{\Lambda}$=0.73, and a velocity-field corrected using the \citet{Mould00} model, which includes the influence of the Virgo cluster, the Great Attractor, and the Shapley supercluster.

\section{Sample selection}
\label{sample}

The sample studied here consists of 24 unobscured type 1 AGN selected from the {\textit{NuSTAR}} public archive. The majority of {\textit{NuSTAR}} observations employed in this work were obtained for detailed broadband spectral analyses published elsewhere (e.g., \citealt{Ballantyne14,Brenneman14,Lohfink15,Zoghbi15,Ursini16}). These observations are typically long (50--150\,ks) and separated into multiple epochs in order to sample spectral variability. The rest of the data were taken as part of the {\textit{NuSTAR}} survey of {\textit{Swift}}/BAT-selected AGN, which consists of short observations (15--25\,ks) of a large sample representative of the local AGN population (Balokovi\'{c} et al., in preparation). Because of the 100-fold increase in sensitivity between {\textit{Swift}}/BAT and {\textit{NuSTAR}}, for any source detected in the {\textit{Swift}}/BAT all-sky survey, even a short exposure results in data with signal-to-noise ratio (SNR) high enough for spectral modeling up to $\sim$70\,keV (e.g., \citealt{Balokovic14,Koss15,Masini16}).

According to the unified model, in type 1 objects we are able to observe the innermost region of the AGN and they are expected to be practically unabsorbed in X-rays. Therefore, we selected all the unobscured type 1 AGN  at low redshifts (z$\leq$0.1) observed with {\textit{NuSTAR}} and with the X-ray data publicly available in the HEASARC\footnote{http://heasarc.gsfc.nasa.gov/} archive. As of April 2016, this sample consists of 67 broad-line AGN. These sources were then cross-correlated with the {\textit{Spitzer}} Heritage Archive (SHA)\footnote{http://irsa.ipac.caltech.edu/applications/Spitzer/SHA} in order to select only those with available {\textit{Spitzer/InfraRed Spectrograph (IRS)}} MIR spectra covering the 5-35~$\mu$m range. The final sample used in this paper comprises 24 objects which are listed in Table \ref{tab1}. We note that we have excluded from this work broad-absorption line (BAL) quasars, because their inner region geometry is thought to be significantly different from standard type 1 AGN. 
\begin{table*}
\centering
\begin{tabular}{lcccccc}
\hline
Name& Optical& R.A.& Dec.& Redshift& Luminosity & Spatial\\
    &type&(J2000)&(J2000)&	  & distance	&scale	\\
    &	&	&	&&(Mpc)	&	(pc arcsec$^{-1}$)\\
\hline
Mrk\,335	& Sy 1.2	&00h06m19.52s&+20d12m10.5s& 0.025&106&490	\\
Fairall\,9	& Sy 1.2		&01h23m45.78s&-58d48m20.8s& 0.047&200&885	\\
Mrk\,1018	& Sy 1.5	&02h06m15.99s&-00d17m29.2s&	0.042&176&787	\\
Mrk\,590	& Sy 1.2	&02h14m33.56s&-00d46m00.1s& 0.026&107&495	\\
Mrk\,1044	& Sy 1		&02h30m05.53s&-08d59m53.3s& 0.016&66&311	\\
3C\,120		& BLRG		&04h33m11.10s&+05d21m15.6s& 0.032&136&620	\\
Ark\,120	& Sy 1		&05h16m11.42s&-00d08m59.4s& 0.032&136&618	\\
1H\,0707-495	& NLSy 1		&07h08m41.50s&-49d33m06.9s& 0.041&174&778	\\
RBS\,0770	& Sy 1.2	&09h23m43.00s&+22d54m32.6s& 0.033&140&637	\\
NGC\,4051	& Sy 1.2	&12h03m09.61s&+44d31m52.8s& 0.003&13&62		\\
NGC\,4151	& Sy 1.5	&12h10m32.58s&+39d24m20.6s& 0.005&20&96		\\
PG\,1211+143	& NLSy 1		&12h14m17.67s&+14d03m13.1s& 0.083&361&1492	\\
NGC\,4593	& Sy 1		&12h39m39.43s&-05d20m39.3s& 0.010&42&198	\\
Mrk\,766	& Sy 1.5	&12h18m26.51s&+29d48m46.3s& 0.015&61&289	\\
MCG-06-30-015	& Sy 1.2	&13h35m53.71s&-34d17m43.9s& 0.007&27&128	\\
IC\,4329A	& Sy 1.2	&13h49m19.27s&-30d18m34.0s& 0.019&80&372	\\
CGCG\,017-073	& Sy 1		&13h49m52.84s&+02d04m45.1s& 0.035&147&666	\\
NGC\,5548	& Sy 1.5 	&14h17m59.53s&+25d08m12.4s& 0.019&80&375	\\
Mrk\,1393	& Sy 1.5	&15h08m53.95s&-00d11m49.0s& 0.056&242&1052	\\
Mrk\,290	& Sy 1.5	&15h35m52.36s&+57d54m09.2s& 0.031&130&593	\\
3C\,382		& BLRG		&18h35m03.39s&+32d41m46.8s& 0.059&253&1094	\\
3C\,390.3	& BLRG		&18h42m08.99s&+79d46m17.1s& 0.057&242&1053	\\
NGC\,7213	& Sy 1.5	&22h09m16.31s&-47d09m59.8s& 0.006&25&120	\\
Mrk\,915	& Sy 1		&22h36m46.50s&-12d32m42.6s& 0.024&101&468	\\

\hline
\end{tabular}						 
\caption{The sample of 24 type 1 AGN studied here sorted by right ascension (R.A.). AGN optical type, R.A. and declination (Dec.) were taken from the NASA/IPAC Extragalactic Database (NED). The redshift, luminosity distance and spatial scale were calculated using a cosmology with H$_0$=73 km~s$^{-1}$~Mpc$^{-1}$, $\Omega_m$=0.27, $\Omega_{\Lambda}$=0.73 and a velocity-field corrected using the \citet{Mould00} model, which includes the influence of the Virgo cluster, the Great Attractor, and the Shapley supercluster.} 
\label{tab1}
\end{table*}

\section{Observations}
\label{observations}

\subsection{X-ray {\textit{NuSTAR}} data}
\label{xray}

High energy X-ray spectra of the sample studied here were observed with the {\textit{NuSTAR}} observatory (angular resolution $\sim$18\arcsec), which consists of two co-aligned hard X-ray telescopes with focal lengths of 10.14\,m. {\textit{NuSTAR}} is the first high energy ($>$10\,keV) orbiting observatory with focusing optics, providing $\sim$2 orders of magnitude increase in sensitivity compared to previous high energy observatories. The data were 
obtained across the 3-80\,keV energy range, using the two {\textit{NuSTAR}} focal planes: the Focal Plane Module A (FPMA) and the Focal Plane Module B (FPMB), which use CdZnTe chips (pixel size of 2.46\arcsec) and have a field of view (FOV) of $\sim$12\arcmin x~12\arcmin ~at 10~keV. A detailed description of the {\textit{NuSTAR}} observatory is given in \citet{Harrison13}.

We processed the raw {\textit{NuSTAR}} data using the standard data processing package, NuSTARDAS, generally following the procedures described in the {\textit{NuSTAR}} user's guide\footnote{http://heasarc.gsfc.nasa.gov/docs/nustar/analysis/nustars$\_$wguide.pdf} and described in more detail in Balokovi\'{c} et al., in preparation. The complete list of observations is given in Table \ref{tab2}. We used a range of software editions (HEASOFT 6.14--6.16, NuSTARDAS 1.4--1.6) and versions of the calibration database (between 20130909 and 20150316). No significant changes were noted in updating from an older to a newer version of the database for any of our targets. Event filtering was performed using the {\textit{nupipeline}} script with the most strict filter setting in order to avoid any possible contamination due to South Atlantic Anomaly (SAA) passages and elevated background levels. Sources exhibited a wide range of variability patterns and amplitudes, but in this work we are concerned only with time-averaged data.

\begin{landscape}
\begin{table}
\scriptsize
\centering
\begin{tabular}{lcccccccc}
\hline
Name		&Obs. & Date&Exp. Time&F$_{3-5~keV}$&F$_{2-10~keV}$ &F$_{7-15~keV}$ &F$_{15-40~keV}$ &F$_{40-80~keV}$\\
		 &ID&(UT)&(ks)&\\
		 
\hline
Mrk\,335	&60001041002&2013/06/13&21.3&(1.35$\pm$0.72)$\times10^{-12}$	&(4.64$\pm$2.01)$\times10^{-12}$	&(2.98$\pm$0.64)$\times10^{-12}$	&(6.63$\pm$0.65)$\times10^{-12}$	&(4.17$\pm$1.05)$\times10^{-12}$\\  
		&60001041003&2013/06/13&21.5&               \\  
		&60001041005&2013/06/25&92.9&               \\
		&80001020002&2014/09/20&68.8&               \\  		
Mrk\,1018	&60160087002&2016/02/10&20.3&(4.91$\pm$0.52)$\times10^{-13}$	&(1.71$\pm$0.11)$\times10^{-12}$	&(1.08$\pm$0.08)$\times10^{-12}$	&(1.62$\pm$0.62)$\times10^{-12}$	&$<$5.04$\times10^{-12}$\\ 
Mrk\,590	&60160095002&2016/02/05&17.9&(8.89$\pm$0.66)$\times10^{-13}$	&(2.96$\pm$0.14)$\times10^{-12}$	&(1.96$\pm$0.10)$\times10^{-12}$	&(3.48$\pm$0.46)$\times10^{-12}$	&$<$1.07$\times10^{-11}$\\
Mrk\,1044	&60160109002&2016/02/08&20.2&(2.73$\pm$0.10)$\times10^{-12}$	&(8.67$\pm$0.23)$\times10^{-12}$	&(4.19$\pm$0.13)$\times10^{-12}$	&(5.98$\pm$0.47)$\times10^{-12}$	&$<$1.50$\times10^{-11}$\\
Fairall\,9	&60001130002&2014/05/09&49.1&(7.30$\pm$0.05)$\times10^{-12}$	&(2.33$\pm$0.16)$\times10^{-11}$	&(1.24$\pm$0.07)$\times10^{-11}$	&(2.04$\pm$0.07)$\times10^{-11}$	&(1.30$\pm$0.13)$\times10^{-11}$\\ 
		&60001130003&2014/05/09&93.6&               \\  
3C\,120		&60001042002&2013/02/06&21.6&(1.60$\pm$0.01)$\times10^{-11}$	&(5.12$\pm$0.11)$\times10^{-11}$	&(2.83$\pm$0.10)$\times10^{-11}$	&(4.57$\pm$0.25)$\times10^{-11}$	&(3.24$\pm$0.09)$\times10^{-11}$\\  
		&60001042003&2013/02/06&127.7&               \\  
Ark\,120	&60001044002&2013/02/18&55.3&(9.99$\pm$3.97)$\times10^{-12}$	&(3.20$\pm$1.19)$\times10^{-11}$	&(1.75$\pm$0.51)$\times10^{-11}$	&(2.97$\pm$0.71)$\times10^{-11}$	&(2.54$\pm$0.11)$\times10^{-11}$\\ 
		&60001044004&2014/03/22&65.3&               \\  
1H\,0707-495	&60001102002&2014/05/05&144.0&(1.90$\pm$0.85)$\times10^{-13}$	&(5.50$\pm$2.52)$\times10^{-13}$	&(1.17$\pm$0.48)$\times10^{-13}$	&(5.04$\pm$4.40)$\times10^{-13}$	&$<$2.59$\times10^{-13}$\\ 
		&60001102004&2014/06/10&48.8&               \\  
		&60001102006&2014/06/28&46.7&               \\  
RBS\,0770	&60061092002&2012/12/26&18.8&(6.88$\pm$0.14)$\times10^{-12}$	&(2.20$\pm$0.03)$\times10^{-11}$	&(1.20$\pm$0.02)$\times10^{-11}$	&(1.92$\pm$0.07)$\times10^{-11}$	&(1.14$\pm$0.39)$\times10^{-11}$\\  
NGC\,4051	&60001050002&2013/06/17&9.4&(5.79$\pm$2.75)$\times10^{-12}$	&(1.87$\pm$0.82)$\times10^{-11}$	&(1.06$\pm$0.32)$\times10^{-11}$	&(2.18$\pm$0.43)$\times10^{-11}$	&(1.34$\pm$0.25)$\times10^{-11}$\\  
		&60001050003&2013/06/17&45.6&               \\  
		&60001050005&2013/10/09&10.2&               \\  
		&60001050006&2013/10/09&49.5&               \\  
		&60001050008&2014/02/16&56.6&               \\  
NGC\,4151	&60001111002&2012/11/12&21.9&(4.08$\pm$0.59)$\times10^{-11}$	&(1.68$\pm$0.18)$\times10^{-10}$	&(1.58$\pm$0.11)$\times10^{-10}$	&(3.37$\pm$0.18)$\times10^{-10}$	&(2.53$\pm$0.27)$\times10^{-10}$\\  
		&60001111003&2012/11/12&57.1&               \\  
		&60001111005&2012/11/14&61.6&               \\  
PG\,1211+143	&60001100002&2014/02/18&111.2&(1.31$\pm$0.26)$\times10^{-12}$	&(4.09$\pm$0.80)$\times10^{-12}$	&(1.99$\pm$0.35)$\times10^{-12}$	&(2.99$\pm$0.25)$\times10^{-12}$	&$<$5.10$\times10^{-12}$\\  
		&60001100004&2014/04/08&48.8&               \\  
		&60001100005&2014/04/09&64.3&               \\ 
		&60001100007&2014/07/07&67.2&               \\ 
NGC\,4593	&60001149002&2014/12/29&23.3&(6.37$\pm$1.82)$\times10^{-12}$	&(2.10$\pm$0.58)$\times10^{-11}$	&(1.26$\pm$0.29)$\times10^{-11}$	&(2.41$\pm$0.44)$\times10^{-11}$	&(2.04$\pm$0.28)$\times10^{-11}$\\ 
		&60001149004&2014/12/31&21.6&               \\ 
		&60001149006&2015/01/02&21.3&               \\ 
		&60001149008&2015/01/04&23.1&               \\ 
		&60001149010&2015/01/06&21.2&               \\ 
Mrk\,766	&60001048002&2015/01/24&81.9&(7.32$\pm$0.07)$\times10^{-12}$	&(2.24$\pm$0.02)$\times10^{-11}$	&(9.95$\pm$0.09)$\times10^{-12}$	&(1.50$\pm$0.03)$\times10^{-11}$	&$<$2.46$\times10^{-11}$\\ 
MCG-06-30-015	&60001047002&2013/01/29&23.3&(1.32$\pm$0.32)$\times10^{-11}$	&(4.18$\pm$0.97)$\times10^{-11}$	&(2.31$\pm$0.35)$\times10^{-11}$	&(3.47$\pm$0.33)$\times10^{-11}$	&(1.70$\pm$0.23)$\times10^{-11}$\\  
		&60001047003&2013/01/30&127.1&               \\  
		&60001047005&2013/02/02&29.6&               \\  
IC\,4329A	&60001045002&2012/08/12&159.0&(3.23$\pm$0.01)$\times10^{-11}$	&(1.07$\pm$0.01)$\times10^{-10}$	&(6.53$\pm$0.02)$\times10^{-11}$	&(1.13$\pm$0.01)$\times10^{-10}$	&(8.61$\pm$0.27)$\times10^{-11}$\\  
CGCG\,017-073	&60160560002&2015/03/31&15.3&(1.99$\pm$0.09)$\times10^{-12}$	&(6.80$\pm$0.21)$\times10^{-12}$	&(4.29$\pm$0.15)$\times10^{-12}$	&(8.39$\pm$0.61)$\times10^{-12}$	&$<$1.98$\times10^{-11}$\\ 
NGC\,5548	&60002044002&2013/07/11&24.1&(9.94$\pm$1.85)$\times10^{-12}$	&(3.50$\pm$0.57)$\times10^{-11}$	&(2.54$\pm$0.27)$\times10^{-11}$	&(4.84$\pm$0.38)$\times10^{-11}$	&(3.82$\pm$0.63)$\times10^{-11}$\\  
		&60002044003&2013/07/12&27.2&               \\  
		&60002044005&2013/07/23&49.4&               \\  
		&60002044006&2013/09/10&51.4&               \\  
		&60002044008&2013/12/20&50.0&               \\  
Mrk\,1393	&60160607002&2016/01/19&21.5&(5.04$\pm$0.44)$\times10^{-13}$	&(1.96$\pm$0.10)$\times10^{-12}$	&(1.63$\pm$0.08)$\times10^{-12}$	&(4.09$\pm$0.50)$\times10^{-12}$	&$<$8.55$\times10^{-12}$\\ 
Mrk\,290	&60061266002&2013/11/14&25.0&(2.25$\pm$0.17)$\times10^{-12}$	&(7.54$\pm$0.38)$\times10^{-12}$	&(4.80$\pm$0.07)$\times10^{-12}$	&(9.30$\pm$0.69)$\times10^{-12}$	&$<$2.43$\times10^{-11}$\\  
		&60061266004&2013/11/27&26.3&     &          \\                     
3C\,382		&60061286002&2012/09/18&16.6&(1.19$\pm$0.47)$\times10^{-11}$	&(3.94$\pm$1.47)$\times10^{-11}$	&(2.30$\pm$0.75)$\times10^{-11}$	&(3.51$\pm$0.05)$\times10^{-11}$	&(2.26$\pm$0.12)$\times10^{-11}$\\  
		&60001084002&2013/12/18&82.4&     &          \\  
3C\,390.3	&60001082002&2013/05/24&23.6&(1.26$\pm$0.01)$\times10^{-11}$	&(4.20$\pm$0.03)$\times10^{-11}$	&(2.52$\pm$0.01)$\times10^{-11}$	&(3.97$\pm$0.05)$\times10^{-11}$	&(2.96$\pm$0.04)$\times10^{-11}$\\  
		&60001082003&2013/05/24&47.5&     &          \\  
NGC\,7213	&60001031002&2014/10/05&101.5&(4.90$\pm$0.05)$\times10^{-12}$	&(1.60$\pm$0.01)$\times10^{-11}$	&(8.66$\pm$0.07)$\times10^{-12}$	&(1.20$\pm$0.02)$\times10^{-11}$	&(7.64$\pm$1.19)$\times10^{-12}$\\  
Mrk\,915	&60002060002&2014/12/02&52.9&(1.72$\pm$0.69)$\times10^{-12}$	&(5.95$\pm$2.33)$\times10^{-12}$	&(4.25$\pm$1.35)$\times10^{-12}$	&(7.52$\pm$2.48)$\times10^{-12}$	&(6.49$\pm$2.80)$\times10^{-12}$\\ 
		&60002060004&2014/12/07&54.1&               \\ 
  		&60002060006&2014/12/12&50.6&               \\ 
\hline
\end{tabular}					 
\caption{Summary of the X-ray observations. Columns 1 to 4 list the object name, the observation ID, the observation date and the exposure time. Columns 5 to 9 correspond to the 3-5, 2-10, 7-15, 15-40 and 40-80~keV band fluxes, respectively. The X-ray fluxes were calculated by averaging all the observations available (the fluxes are given in erg s$^{-1}$ cm$^{-2}$ units). We note that for the individual observations the uncertainties correspond to 1$\sigma$ level, whereas for multiple observations the uncertainties correspond to the standard deviation between the different observations, which are larger than the propagated uncertainties.}
\label{tab2}       
\end{table}
\end{landscape}

We extracted target spectra from circular regions centered on the source in each of the two focal plane modules (FPMA and FPMB), with the radii size chosen according to the total number of counts, between 30\arcsec ~and 150\arcsec . The different sizes used here depend on the brightness of the source. For the faintest objects we used a 30\arcsec ~aperture, whereas for the brightest ones we used larger apertures to collect more photons since the background is low in those cases. The background extraction region is defined as the square area of the detector onto which the target is focused, excluding the circular region 30\% larger than the source extraction region and excluding 30\arcsec ~around any detected serendipitous sources (see \citealt{Lansbury17}). With its spatial resolution of $\sim$28\arcsec ~(half-power diameter; \citealt{Harrison13}), {\textit{NuSTAR}} cannot spatially resolve any of our targets. The source and background spectra were produced together with the ancillary response files using the {\textit{nuproducts}} script. The spectral files were grouped into energy bins so that the median SNR is 5--15 in each bin (depending on the total number of counts), with a minimum SNR of 5 for most sources and 3 for the faintest ones.

\subsection{MIR Spitzer Space Telescope spectra}
\label{spitzer}

We retrieved MIR spectra for the whole sample from the Cornell Atlas of {\textit{Spitzer/IRS}} Source (CASSIS\footnote{http://cassis.astro.cornell.edu/atlas/} v4; \citealt{Lebouteiller11}). The spectra were obtained using the IRS instrument \citep{Houck04}. The bulk of the observations (18/24 galaxies) were made in staring mode using the low-resolution (R$\sim$60-120) IRS modules: the short-low (SL; 5.2-14.5~$\mu$m) and the long-low (LL; 14-38~$\mu$m). The spectra were reduced with the CASSIS software, using the optimal extraction to obtain the best SNR. We only needed to apply a small offset to stitch together the different modules, taking the shorter wavelength module (SL2; 5.2-7.6~$\mu$m) as the basis, which has associated a slit width of 3.6\arcsec.  

We note that for six galaxies (Mrk\,335, Mrk\,1044, 1H\,0707-495, NGC\,4151, NGC\,4593, and Mrk\,766) there are no low-resolution staring mode spectra for the LL module. Therefore, in order to cover the same spectral range, we used the SL low-resolution module together with the high-resolution (HR; R$\sim$600) IRS modules: the short-high (SH; 9.9-19.6~$\mu$m) and the long-high (LH; 18.7-37.2~$\mu$m) available in the CASSIS. The high-resolution module spectra are also reduced with the CASSIS software, using the optimal extraction. We have consistently applied a Gaussian convolution in the HR spectra to degrade the spectral resolution to be the same as that for the low-resolution spectra. The MIR spectra are shown in Appendix \ref{A}. Further details on the {\textit{Spitzer}} observations can be found in Table \ref{tab3}.

As a sanity check, and since we are interested only in AGN-dominated IR data, we estimated the possible contribution from other components to the IRS spectra of our sample. To do so, we used the DeblendIRS\footnote{http://www.denebola.org/ahc/deblendIRS/} routine \citep{Hernan-caballero2015}, that decomposes MIR spectra using a linear combination of three spectral components: AGN, PAH and stellar emission. We found that the MIR spectra of the galaxies in our sample have a high contribution of the AGN ($\sim$90\%; i.e. AGN-dominated systems), as 
expected for type 1 AGN. Further details on the spectral decomposition can be found in Appendix \ref{A}.

\begin{table*}
\centering
\begin{tabular}{lcccccc}
\hline
&\multicolumn{3}{c}{MIR data}&\multicolumn{3}{c}{NIR data}\\
\hline
Name &Telesc./Instr. & Obs. & Date&Telesc./Instr. & Date & Ref.\\
&& ID	& (UT) & or database &(UT)& \\
\hline
Mrk\,335	& {\textit{Spitzer/IRS}} L   & 14448128-1 & 08 Jul 2005	& IRTF/SpeX 	& 25 Jan 2007& a   \\
		& {\textit{Spitzer/IRS}} H   & 14448128-2 & 12 Jan 2009	&$\cdots$&$\cdots$& \\
Fairall\,9	& {\textit{Spitzer/IRS}} L   & 28720896	  & 20 Jan 2009	& 2MASS 	& 21 Oct 1999& b       \\
Mrk\,1018	& {\textit{Spitzer/IRS}} L   & 15076096	  & 28 Jan 2006	& Gemini/GNIRS	& 16 Nov 2012& c   \\
Mrk\,590	& {\textit{Spitzer/IRS}} L   & 18508544	  & 09 Feb 2007	& IRTF/SpeX	& 24 Jan 2007& a   \\
Mrk\,1044	& {\textit{Spitzer/IRS}} L/H & 14447872	  & 04 Aug 2005	& IRTF/SpeX	& 11 Oct 2000& d  \\
3C\,120		& {\textit{Spitzer/IRS}} L   & 18505216	  & 06 Oct 2007	& Gemini/GNIRS	& 15 Dec 2010& e   \\
Ark\,120	& {\textit{Spitzer/IRS}} L   & 18941440	  & 05 Oct 2005	& IRTF/SpeX 	& 26 Jan 2007& a   \\
1H\,0707-495	& {\textit{Spitzer/IRS}} L/H & 14447360	  & 15 Nov 2006	& 2MASS	     	& 26 Feb 2000& b       \\
RBS\,0770	& {\textit{Spitzer/IRS}} L   & 2691392	  & 08 Dec 2004 & 2MASS	     	& 30 Nov 2000& b       \\
NGC\,4051	& {\textit{Spitzer/IRS}} L   & 14449152	  & 10 Dec 2005	& IRTF/SpeX 	& 20 Apr 2002& f   \\
		& {\textit{Spitzer/IRS}} H   & 10342400	  & 01 Jun 2005	&$\cdots$&$\cdots$& \\
NGC\,4151	& {\textit{Spitzer/IRS}} L   & 3754496	  & 08 Jan 2004	& IRTF/SpeX 	& 08 Jan 2006& g   \\
PG\,1211+143	& {\textit{Spitzer/IRS}} L   & 3760896	  & 07 Jan 2004	& Gemini/GNIRS	& 18 Dec 2011& e   \\
NGC\,4593	& {\textit{Spitzer/IRS}} L/H & 4853504	  & 01 Jul 2005	& IRTF/SpeX	& 11 Jun 2006& g   \\
Mrk\,766	& {\textit{Spitzer/IRS}} L/H & 14448896	  & 22 Jun 2006	& IRTF/SpeX	& 21 Apr 2002& f   \\
MCG-06-30-015	& {\textit{Spitzer/IRS}} L   & 4849920	  & 28 Jun 2004	& 2MASS		& 06 Apr 1991& b      \\
IC\,4329A	& {\textit{Spitzer/IRS}} L   & 18506496	  & 29 Jul 2007	& 2MASS		& 25 Jul 1998& b      \\
CGCG\,017-073	& {\textit{Spitzer/IRS}} L   & 26497024	  & 05 Aug 2008	& 2MASS	     	& 28 Feb 2000& b      \\
NGC\,5548	& {\textit{Spitzer/IRS}} L   & 18513152	  & 31 Jul 2007	& IRTF/SpeX 	& 12 Jun 2006& g   \\
Mrk\,1393	& {\textit{Spitzer/IRS}} L   & 15080960	  & 13 Aug 2005	& 2MASS	     	& 11 Mar 1999& b      \\
Mrk\,290	& {\textit{Spitzer/IRS}} L   & 14200320	  & 22 Dec 2005	& IRTF/SpeX 	& 11 Jun 2006& g   \\
3C\,382		& {\textit{Spitzer/IRS}} L   & 11306496	  & 15 Aug 2005	& 2MASS		& 25 Jan 1998& b       \\
3C\,390.3	& {\textit{Spitzer/IRS}} L   & 4673024	  & 24 Mar 2004	& Gemini/GNIRS	& 04 Aug 2011& e   \\
NGC\,7213	& {\textit{Spitzer/IRS}} L   & 18513152	  & 31 Jul 2007	& 2MASS		& 18 Nov 1998& b       \\
Mrk\,915	& {\textit{Spitzer/IRS}} L   & 26495488	  & 29 Jun 2008	& IRTF/SpeX	& 15 Sep 2011& h  \\
\hline
\end{tabular}						 
\caption{Summary of the IR observations. The columns are as follows:
  (1) object name. For the MIR data (2) telescope,
  instrument and grating, where L and H correspond to the low- and
  high-resolution modules, respectively; (3) observation ID; and (4)
  observation date. For the NIR data (5) telescope
  and instrument; (6) observation date; and (7) corresponding references. References: a) \citet{Landt11}; b) \citet{Skrutskie2006}; c) Landt et
  al., in preparation; d) \citet{Rodriguez-Ardila02}; e) \citet{Landt13}; f) \citet{Riffel06}; g) \citet{Landt08}; h) \citet{Lamperti17}.}
\label{tab3}
\end{table*}

\subsection{NIR data}
\label{ground}

In order to complete our IR$-$X-ray correlation analysis we extended our collection of data to include NIR fluxes. To achieve this we compiled cross-dispersed NIR spectra for roughly half of our sample (15/24 sources; see Table \ref{tab2}), obtained either with the SpeX spectrograph \citep{Rayner03} on the 3.0\,m NASA Infrared Telescope Facility (IRTF) or with the Gemini Near-Infrared Spectrograph (GNIRS) on the 8.1\,m Gemini-North Telescope. These spectra are presented in \citet{Rodriguez-Ardila02,Riffel06,Landt08,Landt11,Landt13,Lamperti17} and Landt et al., in preparation. We used the Galactic extinction corrected NIR spectra to derive rest-frame $J$- and $K$-band fluxes measured in a small aperture. 

Finally, for the rest of the sample, we retrieved lower angular resolution $J$- and $K$-band photometry from the Two Micron All-Sky Survey \citep[2MASS;][]{Skrutskie2006} Point Source Catalog\footnote{http://wwww.ipac.caltech.edu/2mass/releases/allsky/}, which were obtained with 1.3\,m telescopes. These fluxes were extracted with the 2MASS point source processing software, using an aperture radius of 4\arcsec. Further details on the NIR observations can be found in Table \ref{tab3}.
 
\section{Spectral modelling}
\label{modelling}

\subsection{X-ray modelling}
\label{xray_modelling}
We performed spectral analysis in {\textit{Xspec}} \citep{Arnaud96}, version 12.8.2. We analysed each observation separately, and fitted spectra from FPMA and FPMB modules simultaneously without coadding. A cross-normalization factor was employed to account for small relative offsets between FPMA and FPMB. In all cases it was found to be consistent with expectations from calibration, typically $<$5\% \citep{Madsen15}. In this work we are only dealing with photometry within the various X-ray bands, and so we did not attempt to separate out different spectral components. Instead, we performed spectrophotometry in the 3-5, 7-15, 15-40, and 40-80\,keV bands. These bands were chosen to avoid the Fe\,K$\alpha$ and sample the continuum and the Compton hump. For each of the bands separately, we assumed a simple power-law model, which provided a satisfactory fit in nearly all cases. For spectra with very high data quality, some of these fits resulted in $\chi^2$ greater than 1.5. In those cases we also fitted the data with a log-parabolic model ({\textit{logpar}} in {\textit{Xspec}}), which provided a better fit and did not 
result in a significantly different band flux. We calculated band fluxes and their statistical uncertainties (defined as 68\% confidence intervals) from these fits based on $\chi^2$ statistics, which are shown in Table \ref{tab2}. These fluxes are corrected for the generally low Galactic column density taken from \citealt{Dickey1990}.

For approximately 40\% of our sample, the data in the 40-80\,keV band is of insufficient quality to yield a robust flux measurement, i.e., there are too few source counts for performing the fit described above. In such cases, we extend the band toward lower energies (down to 30\,keV, or 20\,keV for the faintest sources) and fix the photon index to 2.0 in order to estimate the 40-80\,keV flux. We adopt the upper end of the 68\% confidence interval of these estimates as 1$\sigma$ upper limits and, then, we calculated 3$\sigma$ upper limits which are employed in this study (see Table \ref{tab2}). For completeness and comparison with the literature, we also computed fluxes in the 2-10\,keV band using {\textit{NuSTAR}} data in the 3-5 and 7-15\,keV bands, hence avoiding any flux contributed by the neutral Fe\,K$\alpha$ line detected in many sources in our sample. Fluxes averaged over multiple observation epochs are given in Table \ref{tab2}.

\subsection{MIR modelling}
\label{mir_modelling}

\begin{figure*}
\centering
\par{
\includegraphics[width=8.8cm]{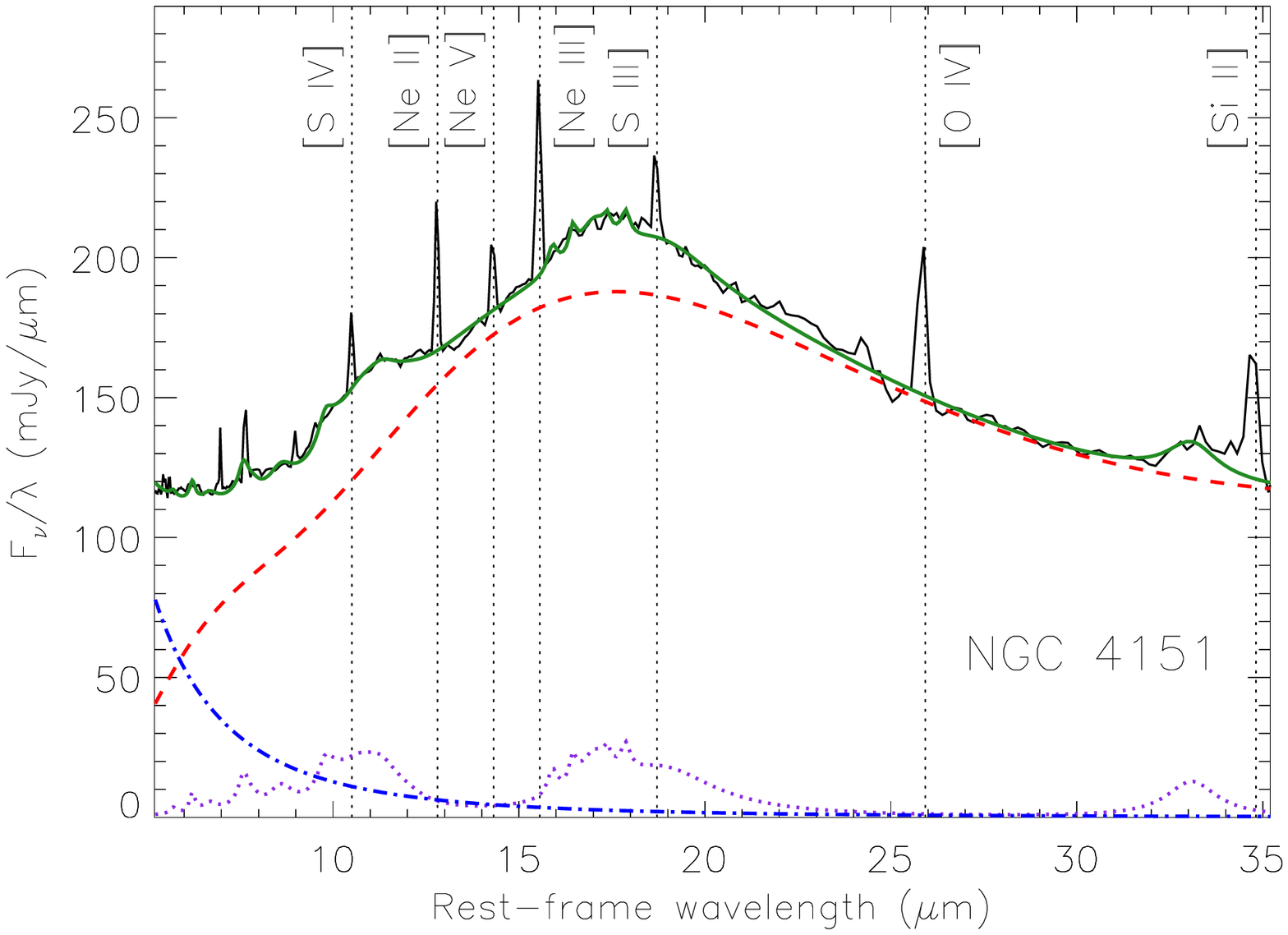}
\includegraphics[width=8.8cm]{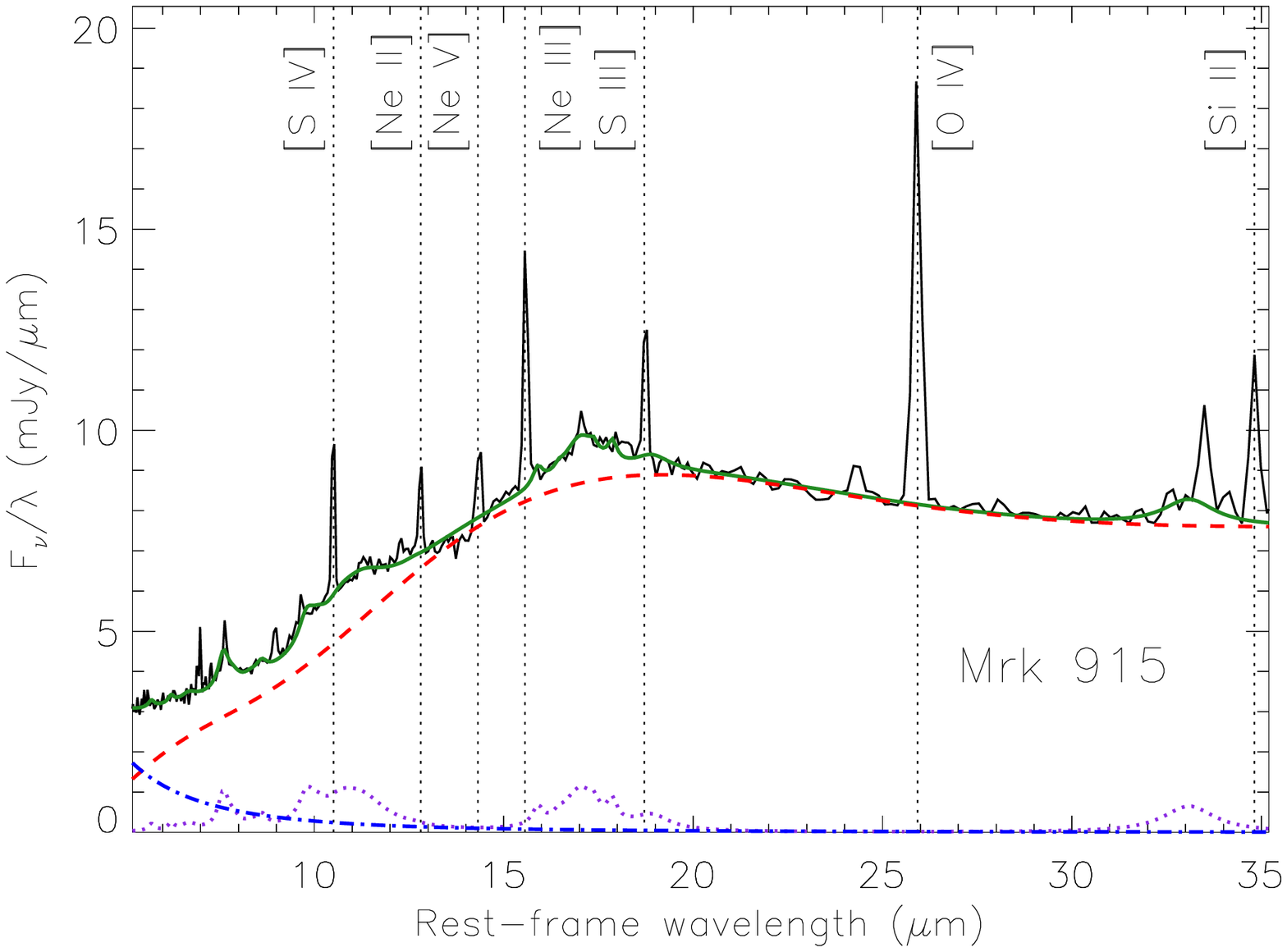}
\par} 
\caption{Examples of the MIR spectral modelling using PAHFIT. We show the {\textit{Spitzer/IRS}} rest-frame spectra (black solid lines), model fits (green solid lines), dust continuum (red dashed lines), dust emission features (purple dotted lines) and stellar continuum (blue dot-dashed lines). The black vertical dotted lines correspond to the MIR emission lines used in this study.}
\label{fig0}
\end{figure*}

We used the PAHFIT v1.2 routine \citep{Smith07} to fit the MIR continuum and the dust emission features of the IRS spectra. The purpose of these fits is to obtain emission-line spectra for all the galaxies. To do so, we used the default model continuum dust temperatures (300, 200, 135, 90, 65, 50, 40 and 35~K) and a stellar continuum component. Fig. \ref{fig0} shows two examples of these fits. 

Once we subtracted the continuum emission and the dust emission features contribution to each MIR spectrum, we obtain the emission-line spectrum for each galaxy. To measure the integrated emission line fluxes we used the DIPSO\footnote{DIPSO is a Starlink package to analyse spectra.} \citep{Howarth04} Emission-Line Fitting routine. Single Gaussians were sufficient to reproduce the line profiles.

In Table \ref{tab4} we list the integrated fluxes of the MIR emission lines used in this study, which are [S\,IV]$\lambda$10.51, [Ne\,II]$\lambda$12.81, [Ne\,V]$\lambda$14.32, [Ne\,III]$\lambda$15.56, [S\,III]$\lambda$18.71,  [O\,IV]$\lambda$25.89 and [Si\,II]$\lambda$34.82 (all wavelengths given in $\mu$m). In the case of the three galaxies for which particular emission lines are not detected we calculated 3$\sigma$ upper limits (see Table \ref{tab4}). 

\begin{landscape}
\begin{table}
\centering
\scriptsize
\begin{tabular}{lcccccccccccc}
\hline
Object&	[S\,IV]$\lambda$10.51&	[Ne\,II]$\lambda$12.81&	[Ne\,V]$\lambda$14.32&	[Ne\,III]$\lambda$15.56&	[S\,III]$\lambda$18.71&	[O\,IV]$\lambda$25.89&	[Si\,II]$\lambda$34.82&J-band&K-band&6~$\mu$m&12~$\mu$m&18~$\mu$m\\
\hline
Mrk\,335&	25.5$\pm$1.8	&	16.6$\pm$1.2&	10.5$\pm$0.2&	25.0$\pm$0.7	&	64.4$\pm$2.8	&	72.9$\pm$1.5	&	78.2$\pm$6.6&	192.0       	&347.6	  	&512.2		&479.7		&441.1	\\
Fairall\,9&	50.3$\pm$7.5	&	33.3$\pm$4.6&  21.5$\pm$6.6&	62.8$\pm$10.1	&	18.8$\pm$2.8	&	70.1$\pm$6.5	&	23.5$\pm$6.5&	231.7       	&332.7   	&806.1		&748.2		&717.5	\\
Mrk\,1018&	9.3$\pm$1.2	&	8.9$\pm$1.3 &   3.4$\pm$1.4&	17.7$\pm$2.4	&	12.6$\pm$2.1	&	24.6$\pm$2.7	&	15.2$\pm$2.3&	91.0	 	& 84.8	  	&182.0		&141.9		&128.9	\\
Mrk\,590&	20.8$\pm$5.1	&	27.4$\pm$1.1&   7.4$\pm$3.1&	31.2$\pm$4.2	&	10.9$\pm$1.7	&	32.9$\pm$2.5	&	29.4$\pm$9.5&	359.8 		&291.9   	&112.4		&234.0		&309.1	\\
Mrk\,1044&	19.3$\pm$3.4	&	45.7$\pm$1.6&	22.0$\pm$0.3&	28.5$\pm$0.6	&	16.5$\pm$0.7	&	16.4$\pm$0.3	&	65.4$\pm$3.1&	136.7    	&272.3   	&304.0		&270.5		&237.5	\\
3C\,120	&	202.7$\pm$9.7	&	49.9$\pm$4.1&173.3$\pm$11.3& 268.3$\pm$9.2	&	50.7$\pm$13.0	&	875.5$\pm$21.5&	142.0$\pm$16.2&		178.7      	&383.2   	&528.2		&640.8		&759.2	\\
Ark\,120&	44.5$\pm$15.5	&    	26.3$\pm$4.0&  15.7$\pm$1.4&  37.1$\pm$5.4	&	14.5$\pm$3.1	&	60.6$\pm$6.6	&	33.2$\pm$8.4&	717.5         	&1053.8   	&798.8		&627.0		&525.0	\\
1H\,0707-495&	6.3$\pm$0.9	&	3.8$\pm$0.2 &	5.0$\pm$0.1 &	21.3$\pm$0.6	&	$<$5.5		&	21.1$\pm$1.0	&	32.2$\pm$3.3&	64.4       	& 74.4    	&57.1		&65.1		&70.0 	\\
RBS\,0770&	36.5$\pm$7.4	&	25.0$\pm$1.9&	23.4$\pm$2.0&	70.1$\pm$2.3	&	20.5$\pm$1.7	&	70.8$\pm$3.4	&	40.0$\pm$5.4& 	240.1       	&271.4   	&261.5		&289.9		&308.3	\\
NGC\,4051&	46.3$\pm$17.6	&      198.5$\pm$5.1& 126.4$\pm$5.0&	214.2$\pm$1.9	&	107.5$\pm$4.0	&	354.0$\pm$3.6	&	122.2$\pm$6.6&	430.8       	&559.4  	&645.9		&1259.2		&1415.6	\\
NGC\,4151&	957.9$\pm$33.7	&    1606.3$\pm$42.8&796.1$\pm$58.3&	2352.0$\pm$50.6&854.73$\pm$61.5& 1990.0$\pm$101.4	&	1404.6$\pm$192.1&    	1084.2         	&1683.4  	&3503.0		&4916.4		&6410.4	\\
PG\,1211+143&	$<$32.3	&	6.2$\pm$0.6 &   4.2$\pm$0.6&	11.9$\pm$1.9	&	$<$9.6		&	9.8$\pm$3.0	&	$<$22.7&		127.0       	&202.9   	&391.9		&402.4		&403.4	\\
NGC\,4593&	33.3$\pm$17.5	&	61.2$\pm$3.9&	77.3$\pm$3.1&	115.9$\pm$3.7	&	50.8$\pm$0.7 	&	112.2$\pm$2.2	&	102.1$\pm$3.1&	393.5         	&707.3   	&971.9		&846.4		&841.1	\\
Mrk\,766&	93.4$\pm$30.3	&     225.4$\pm$10.7& 229.7$\pm$2.0&	290.0$\pm$2.2	&	125.6$\pm$4.0	&	454.5$\pm$1.5	&	114.6$\pm$3.9&	285.0         	&517.3   	&502.7		&825.2		&1332.4	\\
MCG-06-30-015&	85.6$\pm$13.6	&	48.0$\pm$7.0& 40.9$\pm$11.4&	60.0$\pm$4.4	&	56.6$\pm$5.8	&	140.4$\pm$14.6&	31.9$\pm$5.9& 		506.8       	&711.4   	&499.9		&596.8		&675.6	\\
IC\,4329A&	384.4$\pm$45.0	&     234.8$\pm$14.3&351.1$\pm$19.9&	611.6$\pm$37.6&225.4$\pm$29.3&993.4$\pm$44.2	&	434.9$\pm$46.3&      		1159.9         	&1749.4  	&2561.7		&2672.8		&2868.2	\\
CGCG\,017-073&	12.5$\pm$1.1	&	9.1$\pm$1.8 &	14.0$\pm$4.4&	20.1$\pm$2.0	&	43.0$\pm$1.7	&	42.3$\pm$1.4	&	18.4$\pm$3.1&   131.6         	&163.6   	&74.1		&89.5		&99.9	\\
NGC\,5548&	59.1$\pm$3.4	&	88.5$\pm$3.9&	33.4$\pm$9.1&	93.0$\pm$7.4	&	43.2$\pm$5.5	&	89.1$\pm$15.5	&	97.8$\pm$24.5&	277.6       	&364.4   	&321.1		&508.2		&652.4	\\
Mrk\,1393&	37.8$\pm$1.2	&	15.5$\pm$1.1&	28.0$\pm$2.3&	63.5$\pm$5.2	&	21.7$\pm$1.8	&	116.5$\pm$3.2	&	13.3$\pm$3.6&	101.6        	& 97.1    	&64.5		&88.7		&140.1	\\
Mrk\,290&	24.1$\pm$1.5	&	23.0$\pm$1.2&	20.9$\pm$2.7&	42.5$\pm$2.3	&	17.7$\pm$4.4	&	56.6$\pm$5.1	&	11.9$\pm$3.8&	167.5       	&233.2   	&173.5		&225.7		&272.3	\\
3C\,382	&	11.9$\pm$1.0	&	10.6$\pm$1.2&	 7.3$\pm$0.4&	19.9$\pm$3.4	&	8.8$\pm$1.3	&	18.2$\pm$1.6	&	19.1$\pm$2.2&	407.8       	&497.2   	&404.8		&226.1		&182.6	\\
3C\,390.3&	10.3$\pm$2.2	&	35.3$\pm$1.3&	 7.6$\pm$2.2&	31.4$\pm$2.4	&	8.5$\pm$0.7	&	31.8$\pm$3.3	&	18.4$\pm$2.9&	153.9      	&262.1   	&252.4		&295.3		&346.9	\\
NGC\,7213&	$<$225.3	&      258.0$\pm$6.1&	10.4$\pm$2.0&	162.0$\pm$7.7	&	26.4$\pm$7.8	&	18.0$\pm$1.5	&	138.4$\pm$14.3&1677.1       	&1486.3  	&456.1		&668.6		&680.3	\\
Mrk\,915&	137.4$\pm$4.6	&	58.6$\pm$4.7&	68.8$\pm$11.3&	170.8$\pm$2.9	&	82.1$\pm$5.2	&	344.1$\pm$4.5	&	96.9$\pm$8.1&	160.6       	&196.3   	&102.8		&205.1		&292.9	\\
\hline
\end{tabular}						 
\caption{MIR emission line fluxes (in 10$^{-15}$~erg~s$^{-1}$~cm$^{-2}$ units) and $J$-, $K$-, $N$- and $Q$-band fluxes (in 10$^{-13}$~erg~s$^{-1}$~cm$^{-2}$ units) employed here. The estimated uncertainties for the $J$- and $K$-band (1.2 and 2.2~$\mu$m), and MIR bands (6, 12 and 18~$\mu$m) are $\sim$10\% and $\sim$20\%, respectively.} 
\label{tab4}
\end{table}
\end{landscape}

\section{The Correlation Spectrum Technique (CST)}
\label{correlation_spectrum}

As explained in Section \ref{intro}, hard X-rays are good tracers of the AGN power due to the high energies involved in the accretion process. On the other hand, dust in the central region of AGN reprocesses some fraction of these high-energy photons, which are then re-emitted in the IR. For this reason the MIR--X-ray correlation has been widely studied for different types of AGN at various MIR wavelengths ($\sim$6 to 25~$\mu$m; see \citealt{Bernete16} and references therein). However, to date there have been no detailed studies across the entire IR spectral range, and in relation to a set of X-ray bands defined from 3 to 80~keV. In order to do this we employ a new technique which can be used to determine the IR wavelength at which the correlation peaks and to identify specific features closely linked to the AGN.

\begin{figure*}
\centering
\par{
\includegraphics[width=8.812cm]{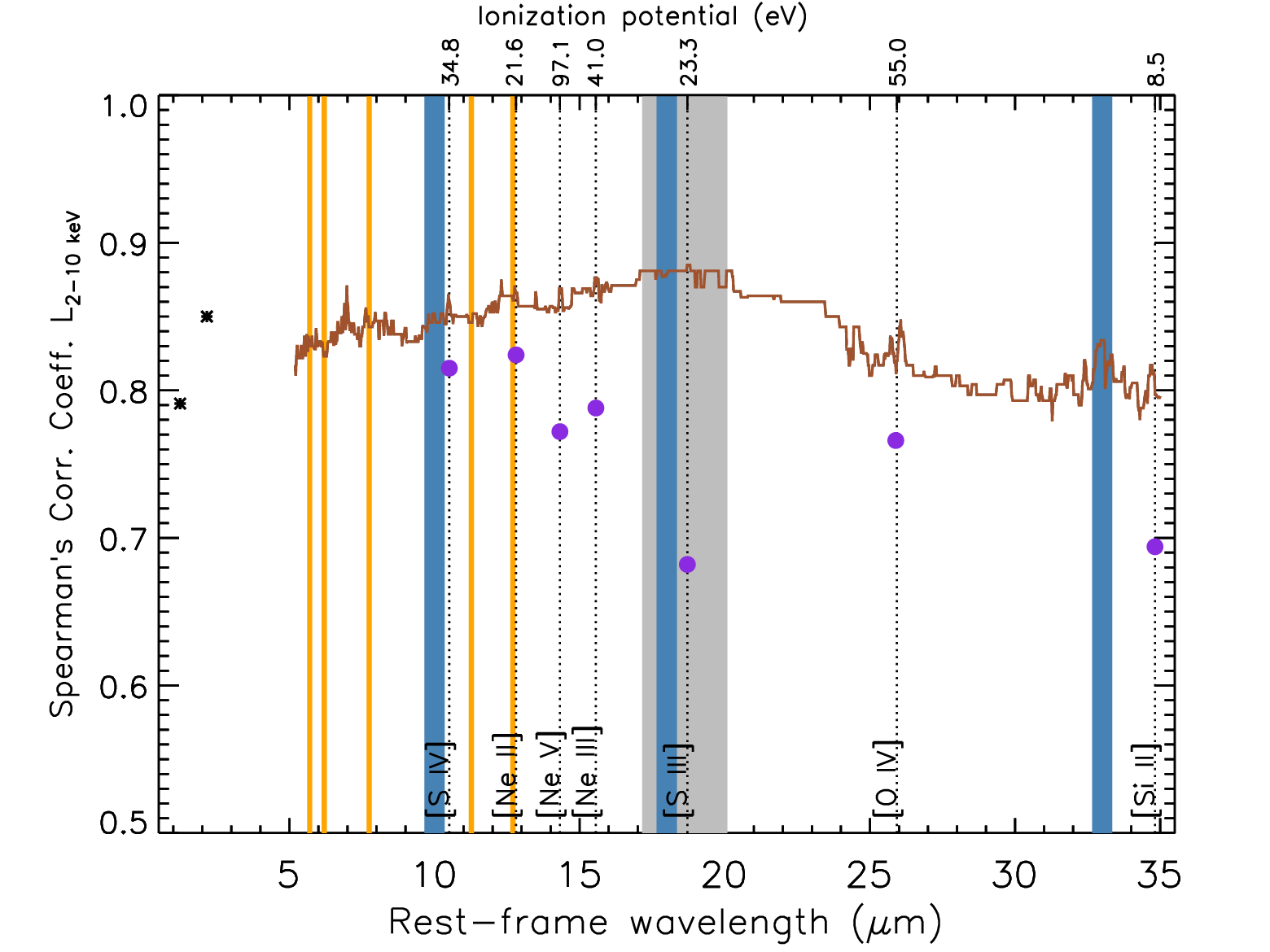}
\includegraphics[width=8.812cm]{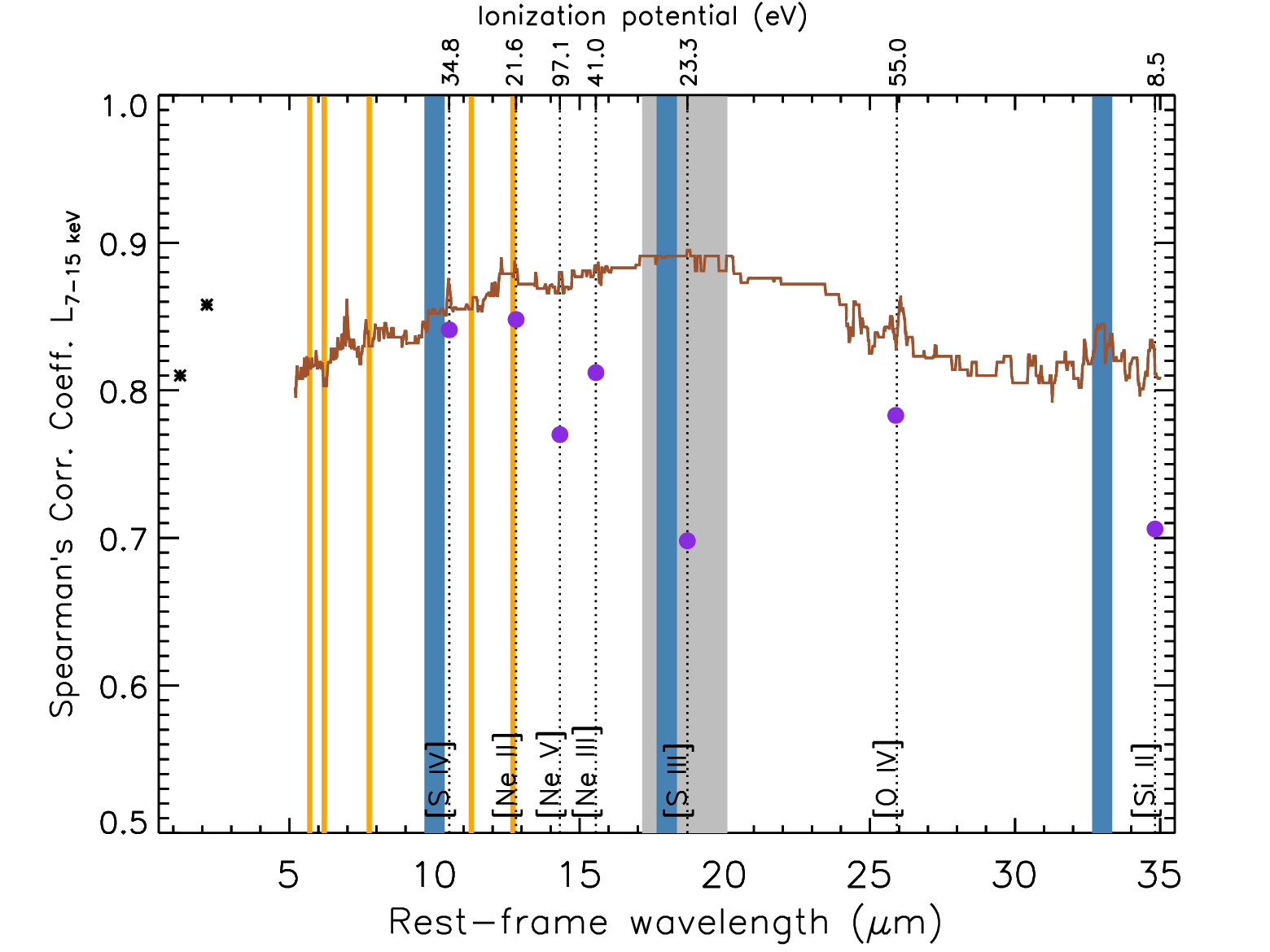}
\includegraphics[width=8.812cm]{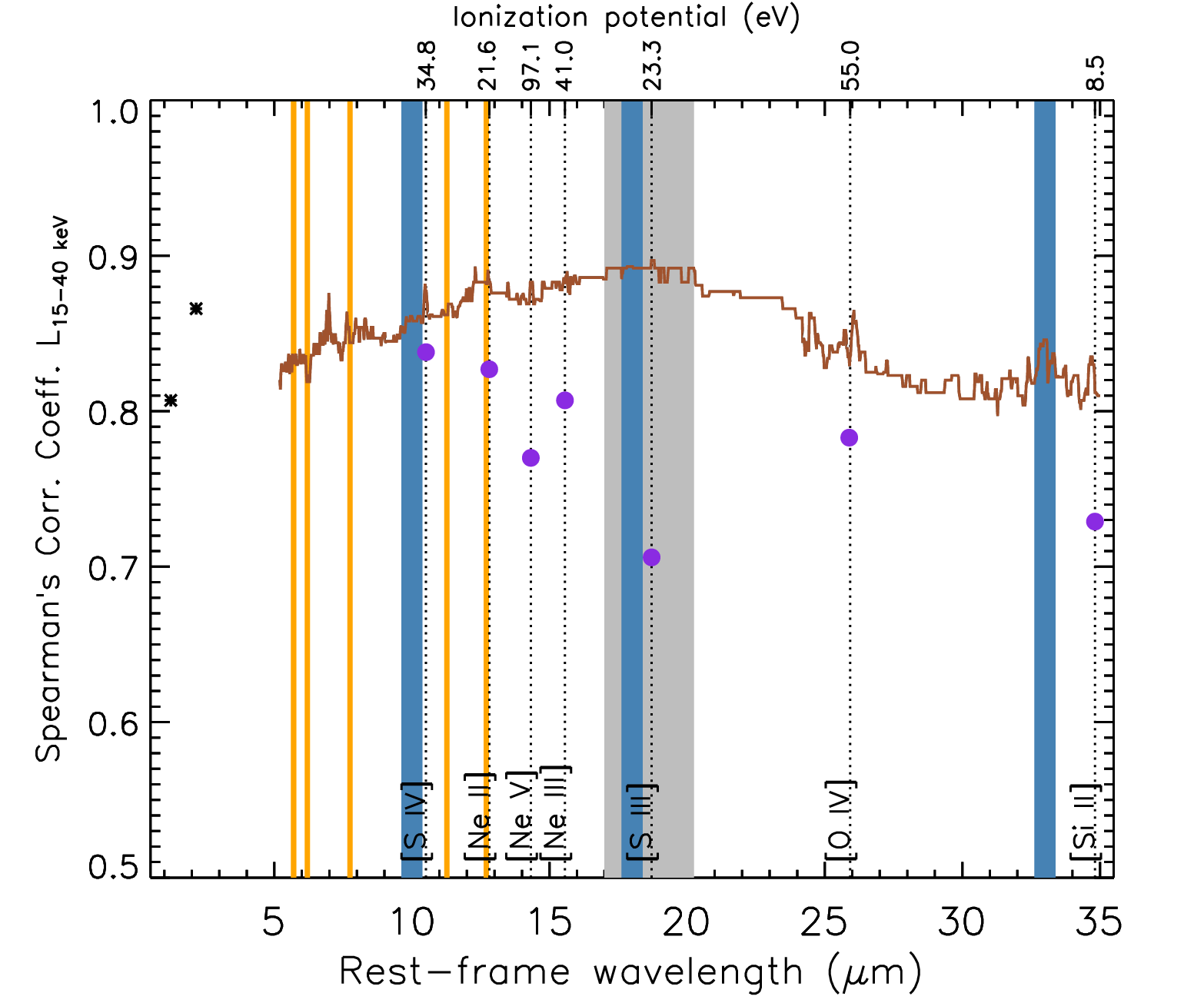}
\includegraphics[width=8.812cm]{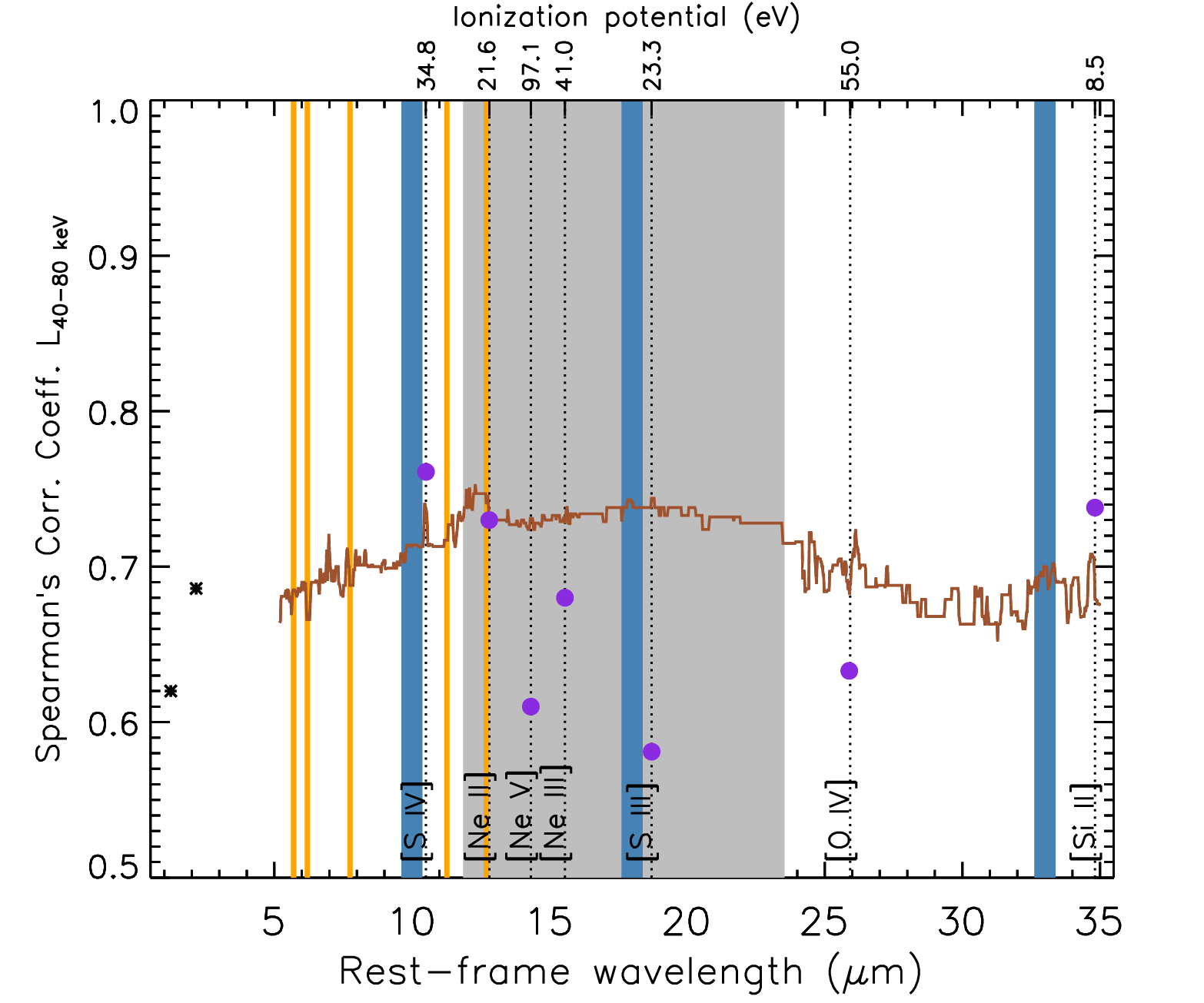}
\par} 
\caption{Correlation spectra of the sample of 24 type 1 AGN studied here. From top left to bottom right panels: 2-10, 7-15, 15-40 and 40-80~keV bands. We show the rest-frame IRXCS (brown solid line), the NIR (black asterisks) and the MIR emission line correlation coefficients (purple circles). The orange and blue vertical solid lines correspond to PAH and silicate features, respectively, and the black vertical dotted lines are the MIR emission lines. The grey vertical region identifies the maximum of the correlation.} 
\label{fig1}
\end{figure*}

Here we present the IR to X-ray correlation spectrum (hereafter IRXCS) for our sample of 24 type 1 AGN. We used the correlation spectrum technique (CST), developed for medium to large samples of objects by \citet{Jin12}. This technique consists of cross-correlating the monochromatic luminosities at each wavelength of the entire rest-frame spectrum with a given X-ray band for a sample of sources. First, we used the \textit{Spitzer/IRS} MIR spectral range ($\sim$5-35~$\mu$m) and calculated the monochromatic luminosity at $\sim$1000 wavelengths evenly distributed throughout it. Then we used the Spearman's rank test to evaluate the correlation coefficients at each of these monochromatic MIR luminosities with all the X-ray bands considered here, resulting in the IRXCS for each band. We then plot the correlation coefficients for the 2-10, 7-15, 15-40 and 40-80~keV bands\footnote{Note that we do not show the 3-5~keV IRXCS in the Fig. \ref{fig1} since this X-ray band is included in the 2-10~keV band and the resulting correlation spectra are practically identical.} against wavelength as brown solid lines in Fig. \ref{fig1}. The diagnostic power of this method is that it can reveal how a general correlation between a spectral range, in this case the IR, changes across it. We refer the reader to \citet{Jin12} for further details on the CST technique.

In addition to the MIR correlation spectra, in Fig. \ref{fig1} we included the NIR correlation coefficients ($J$- and $K$-bands; black asterisks) to increase the IR spectral coverage, and also those measured for the MIR emission lines (purples circles). Finally, we list the IPs of the MIR emission lines (black vertical dotted lines), the PAH and silicate features (orange and blue vertical solid lines), and the maximum of each IRXCS (grey vertical region).

The correlation coefficients plotted in Fig. \ref{fig1} were calculated taking into account the upper limits 
to the MIR emission line luminosities (see Table \ref{tab4}) and also to the 40-80~keV band luminosities (see Table \ref{tab2}). In order to do that, the correlations were carried out with the ASURV survival analysis package \citep{Feigelson85,Isobe86} using the Spearman's rank test and the Bucley-James method for censored data.

To test the survival analysis we first repeated the correlations for the emission lines without including the upper limits. As expected, we found practically the same results because there are few upper limits (only three galaxies and five upper limits in total). In contrast, the correlation spectrum when we do not consider the upper limits to the 40-80~keV band luminosities is significantly different. This indicates that in this case the number of upper limits is too large (40\% of the sample) to yield a robust IRXCS. 

We then produced a new version of Fig. \ref{fig1} for the 15 sources without upper limits in the 40-80~keV band 
(see Appendix \ref{B}). We found the same correlation strengths and shape for all the X-ray energy bands, confirming that the lower correlation coefficients shown for the 40-80~keV band IRXCS in the bottom-right panel of Fig. \ref{fig1} are just a consequence of including a large number of upper limits and so do not correspond to any real trend. Therefore in the following we will only consider the results for the 3-5, 2-10, 7-15 and 15-40~keV bands, although we note that the spectral shape of the 40-80~keV IRXCS is the same as those of the lower energy bands.

Finally, we note that the luminosity-luminosity correlations might be caused, at least in part, by distance effects. Therefore, we also checked the same correlations in flux-flux space. The correlations using the luminosities are stronger than the flux-flux correlations, but the latter are still significant and show practically the same trends.

\section{The IRXCS of type 1 AGN}
\label{results}

From the analysis performed here, we find that all the hard X-ray bands correlate well with the entire IR spectrum, with the correlation maximum being around $\sim$15-20~$\mu$m (see Fig. \ref{fig1}). In general the correlation spectra are smooth, although they show some features. In the MIR, the most prominent are the silicate band at $\sim$10-10.5~$\mu$m, the $\sim$33-34~$\mu$m crystaline silicate broad plateau (e.g., \citealt{Sturm00,Smith07}) and various MIR emission lines. See Section \ref{Discussion} for discussion of the implications of these findings.

In the NIR, we find a stronger correlation for the K-band than for the J-band. While the J-band data follows the trend of the bluest part of the MIR correlation (correlation coefficient $\sim$0.80), the K-band data correlates with the X-rays at a similar level as found for the 10~$\mu$m emission ($\rho\sim$0.85). The result is a correlation peak at $\sim$2~$\mu$m. Although for the NIR data we have not confirmed that all the fluxes are AGN-dominated, as we did for the MIR spectra (see Appendix \ref{A}), the apertures used to extract the NIR fluxes are similar or smaller than the IRS aperture. Therefore we are confident that for the majority of the galaxies the NIR fluxes are indeed AGN-dominated. Note that we find a less significant correlation for the K-band luminosities ($\rho\sim$0.77) and no changes for the J-band luminosities by using the lower angular resolution 2MASS fluxes for all the sample. This indicates that by using higher angular resolution fluxes we can recover AGN-dominated fluxes in the K-band, but not in the J-band, where the host galaxy emission is much higher, as claimed by \citet{kotilainen92}. These authors found that the non-stellar fraction generally increases towards longer NIR wavelengths in a complete hard X-ray selected sample of AGN (mainly type 1). Indeed, they found that for the majority of the sources in their sample the K-band continuum from an aperture of 3\arcsec ~diameter is dominated by non-stellar emission. 

\begin{figure*}
\centering
\par{
\includegraphics[width=8.7cm]{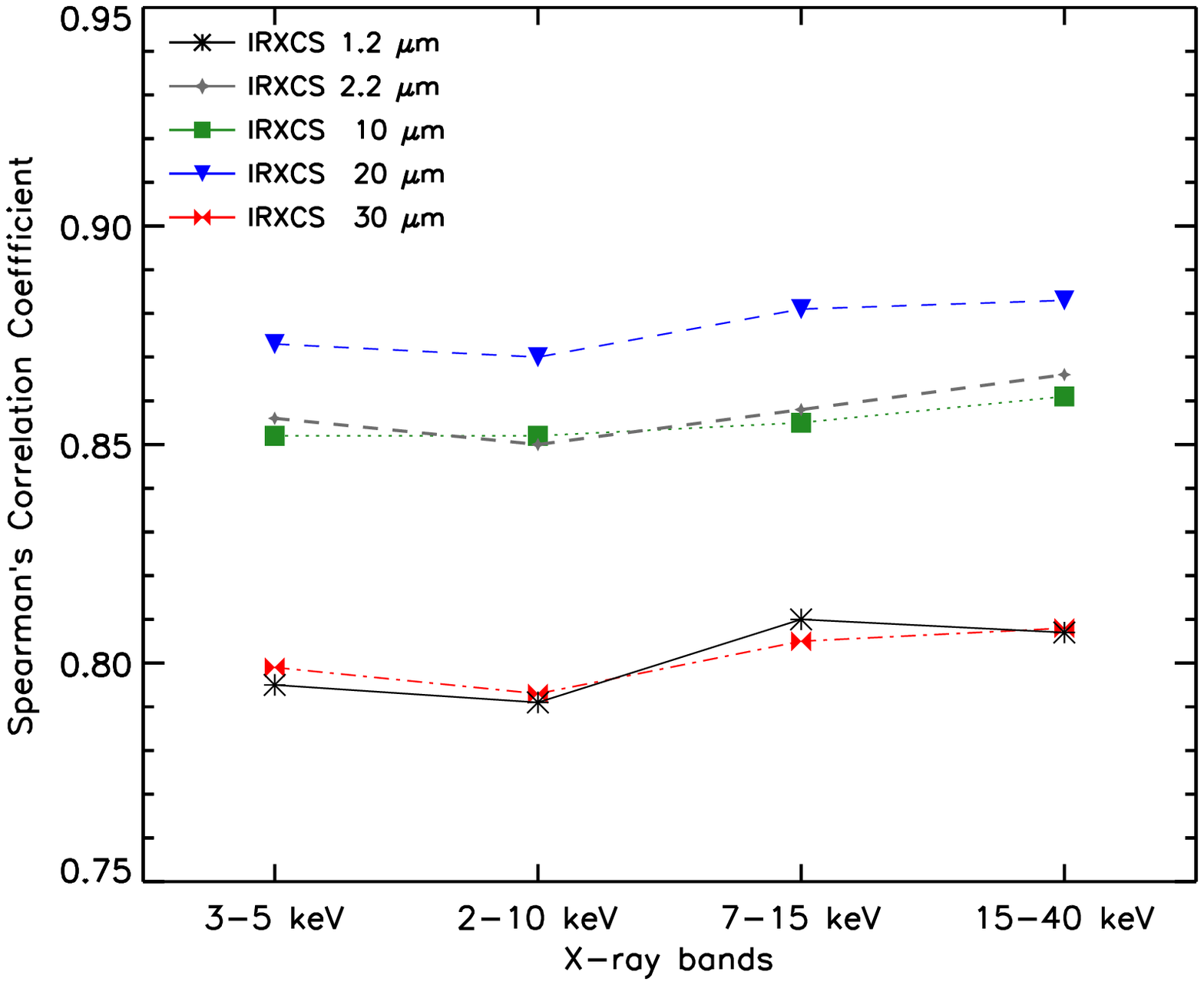}
\includegraphics[width=8.7cm]{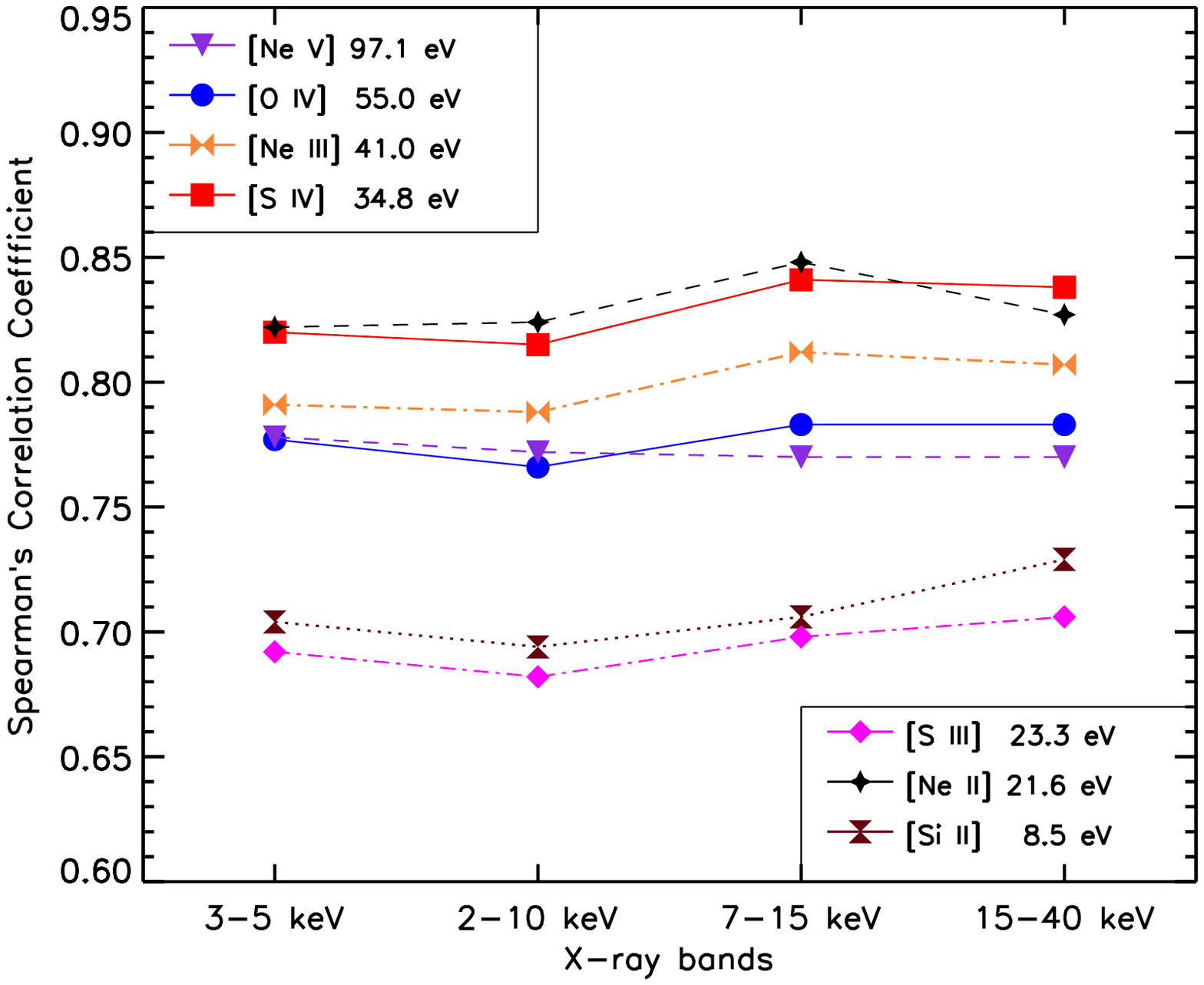}
\par} 
\caption{Comparison between the different correlation coefficients for each X-ray band. Left panel: Correlation coefficients at different continuum wavelengths (1.2, 2.2, 10, 20 and 30~$\mu$m). Right panel: same as in the left panel, but for the MIR emission lines.}
\label{fig2}
\end{figure*}

By comparing the results for the 3-5, 2-10, 7-15 and 15-40~keV bands, we observe that the correlation coefficients derived here are practically identical (see left panel of Fig. \ref{fig2}). The correlation coefficients are $\sim$0.87-0.88 at 20~$\mu$m, $\sim$0.85-0.86 at 2.2 and 10~$\mu$m, and $\sim$0.80 at 1.2 and 30~$\mu$m.

\begin{figure*}
\centering
\includegraphics[width=19.00cm]{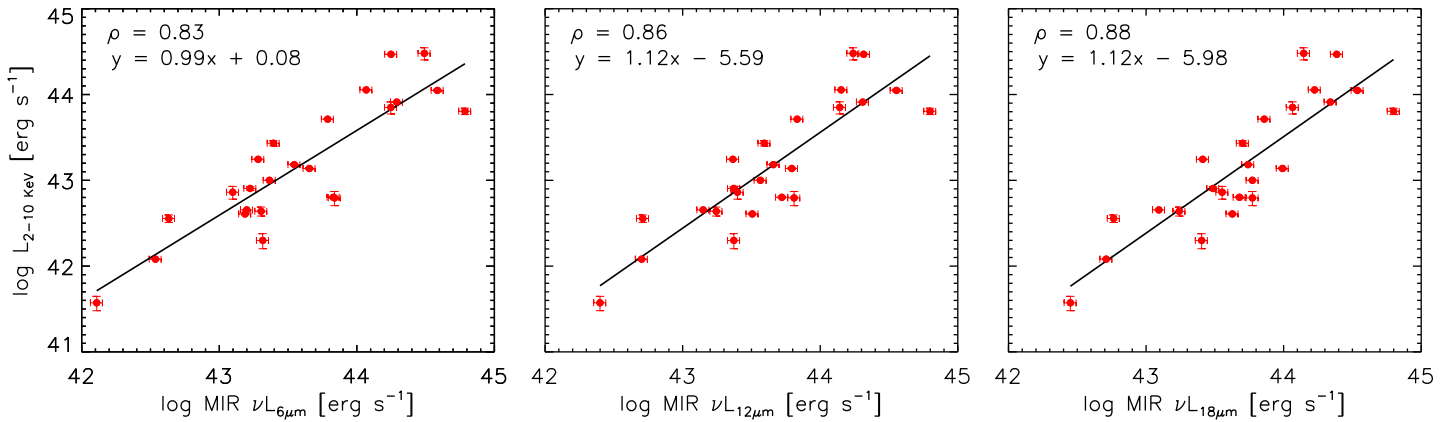}
\caption{MIR$-$2-10~keV X-ray luminosity correlations (using 6, 12 and 18~$\mu$m luminosities). The vertical error bars correspond to the 2-10~keV uncertainties listed in Table \ref{tab2}. The horizontal error bars correspond to the estimated uncertainty of the MIR fluxes, which is $\sim$20\% for the {\textit{Spitzer}} spectra.}
\label{fig3}
\end{figure*}

We also performed linear regressions for the commonly used 2-10~keV X-ray band in log-log space using the Bucley-James least squares method (see \citealt{Feigelson85} and \citealt{Isobe86} for further details) to compare with previous studies. In Fig. \ref{fig3} we show these linear regressions using three monochromatic luminosities measured from the MIR spectra (6, 12 and 18~$\mu$m). We find tight correlations for the three MIR luminosities considered ($\rho=$ 0.83, 0.86 and 0.88, respectively; see Fig. \ref{fig3}). This in agreement with the results reported by \citet{Asmus2015} using the same X-ray band and high angular resolution MIR data at 12 and 18 $\mu$m for samples of 152 and 38 AGN, respectively. On the other hand, using lower angular resolution MIR data from the {\textit{Infrared Imaging Satellite ``AKARI''}}, \citet{Matsuta12} reported a higher correlation significance between the 14--195~keV and 9~$\mu$m luminosities than for those at 18~$\mu$m for a large sample of type 1 AGN.

The slopes that we measure for the MIR--2-10~keV correlations are $\sim$1 (see Fig. \ref{fig3}), in agreement with previous studies using the same X-ray band and similar MIR wavelengths ($\sim$6 to 25~$\mu$m; see \citealt{Bernete16} and references therein). For example, using data from {\textit{ISOCAM}} and the {\textit{Wide-field Infrared Survey Explorer (WISE)}} at $\sim$6~$\mu$m, \citet{Ramos07} and \citet{Mateos15} found slopes of $\sim$1.0 and $\sim$1.1 for type 1 AGN respectively. On the other hand, \citet{Fiore09} reported a flatter slope of $\sim$0.7 by using 5.8~$\mu$m fluxes of type 1 AGN from the {\textit{Spitzer/Infrared Array Camera (IRAC)}} with log~$L_{(5.8~\mu m)}>43.04$. Using high angular resolution 12~$\mu$m fluxes from VISIR, \citet{Gandhi09} found a slope of $\sim$1.2 for a sample of type 1 AGN.

We also find identical slopes (1.12) at 12 and 18~$\mu$m (see central and right panels of Fig. \ref{fig3}), which is in agreement with the results reported by \citet{Asmus2015} and \citet{Ichikawa16}. \citet{Asmus2015} found slopes of $\sim$1.0 by using high angular resolution fluxes at 12 and 18~$\mu$m for large samples of AGN, and \citet{Ichikawa16} reported the same slopes for the 12 $\mu$m--14-195~keV and 22 $\mu$m--14-195~keV correlations (1.04 and 1.02 respectively) using low angular resolution MIR data of a large sample of AGN.

Regarding the MIR emission lines (shown as purple circles in Fig. \ref{fig1}), we do not find a clear correlation between IP and the correlation strength. This is shown in the right panel of Fig. \ref{fig2}. All the lines considered here show correlation coefficients $\geq0.7$. It is worth noting the strong correlation between the low IP [Ne\,II] line and the X-rays ($\rho\sim$0.83). The strength of the correlation is practically the same for the [Ne\,II] and [S\,IV] lines, closely followed by [Ne\,III], [O\,IV] and [Ne\,V]. Finally, the lower correlation coefficients are those of [S\,III] and [Si\,II].

As for the IR continuum, the emission line correlation coefficients are practically constant in the four X-ray bands considered (see right panel of Fig. \ref{fig2}). 

\section{Discussion}
\label{Discussion}

As explained in Section \ref{sample}, unobscured type 1 AGN are ideal laboratories for studies of the emission from, and components within, the nuclear region. Indeed we confirm that the MIR spectra of our sample galaxies are AGN-dominated ($\sim$90\% of AGN contribution; see Appendix \ref{A}). This result, together with the tightness of the correlations between the hard X-rays and the entire NIR-to-MIR range (see Fig. \ref{fig1}) confirms that the IR emission of our sample is AGN-dominated. 
  
The correlation method applied here allows us to examine how the IR versus X-ray correlations change across the entire IR spectral coverage ($\sim$1--35~$\mu$m). The IRXCS for various hard X-ray energy bands between 3 and 40 keV shows a correlation maximum at $\sim$15-20~$\mu$m (see Fig. \ref{fig1}). This correlation peak coincides with the maximum AGN contribution to the MIR spectra for the majority of the sources analyzed here (see Appendix \ref{A}). At $\lambda\gtrsim$20~$\mu$m the correlation spectrum slowly decreases. 

Although the spectral shape of the IRXCS are relatively smooth, there are some specific features. An example is the silicate feature, which for our sample peaks at  $\sim$10-10.5~$\mu$m (see Fig. \ref{fig1} and Appendix \ref{A}), as generally found for type 1 AGN \citep{Hao05,Landt10,Hatziminaoglou15}. Another feature in the correlation spectrum is the $\sim$33-34 $\mu$m broad plateau (e.g., \citealt{Sturm00,Smith07}). This feature is generally attributed to crystalline silicates such as olivines and it has also been detected in the MIR spectra of comets and planetary nebulae \citep{Koike93,Waters98}. The significant correlation of this emission feature with the hard X-rays indicates that it is produced by dust heated by the AGN. A further investigation on the influence of spectral features would be to check whether the PAH features, which are primarily related to star formation, show a de-correlation. However, the galaxies in our sample show weak or absent PAH lines (fractional contribution $<$10\% according to our spectral decomposition analysis; see Table \ref{tab_appendix}), preventing us from performing this test. From Fig. \ref{fig1} it seems that the 6.2~$\mu$m band is de-correlated with the X-rays, but the others are not. A sample of AGN with strong PAHs is needed to confirm this de-correlation. Another interesting result is that we observe a tendency in the different IR features detected in the IRXCS to be more prominent when they are far from the correlation maximum. This could be the result of dilution of the features where the AGN continuum contribution is at its maximum.

In the NIR we see the correlation peak at $\sim$2~$\mu$m, as the K-band emission is more strongly correlated with the X-rays than the J-band emission. This could be related to the NIR bump observed in the spectral energy distributions (SEDs) of some type 1 AGN ($\sim$1-5~$\mu$m; \citealt{Mor09,Herrero11,Mor12,honig13,Mateos16,Hernan-caballero2016}). Various origins have been proposed for this NIR bump: 1) an extra-contribution to the nuclear fluxes from the host galaxy; 2) emission from a compact disc of hot dust detected in interferometric MIR data of some Seyfert galaxies (e.g., \citealt{honig13,Tristram14}); and 3) emission from a hot pure-graphite component located in the outer part of the BLR, as proposed by \citet{Mor09}.

Regarding the latter possibility, in \citet{Mor09} the authors found for a sample of 26 type 1 quasars that their NIR SEDs required an extra hot dust component (T$\sim$800-1800~K), in addition to a clumpy torus model, to be reproduced. The widely used clumpy torus models of \citet{Nenkova08,bNenkova08} assume a standard Galactic grain composition of 53\% silicates and 47\% graphites, and the sublimation temperature of silicate grains ($\sim$1500~K). \citet{Mor12} proposed that to reproduce the NIR bump, a collection of dusty clouds of pure-graphite (T$_{sub}\sim$1800~K) in the outer part of the BLR is needed. This difference between the sublimation temperature of silicates and graphites has been accounted for in the latter version of the CAT3D torus models \citep{honig10}, and it explains a wider variety of SED shapes (Garc\'ia-Gonz\'alez et al. in preparation). On the other hand, using MIR interferometry data of the Seyfert 1 galaxy NGC\,3783, \citet{honig13} proposed a different scenario to reproduce the nuclear IR SED of Seyfert galaxies. The MIR bump ($\sim$15--20~$\mu$m) would be produced by polar dust within the inner parsecs of the galaxy and the NIR bump ($\sim$1-5~$\mu$m) by a compact disc structure (i.e. the torus). The results presented here indicate that this NIR bump is strongly correlated with the hard X-rays, confirming that it is AGN-related, but we cannot favour either of the two alternative origins discussed above.

The broad energy coverage of the {\textit{NuSTAR}} data studied here allows us to divide the X-ray continuum emission in a set of bands between 3 and 80 keV. We find that both the shape and the correlation strengths of the IRXCS are independent of the X-ray band chosen (see left panel of Fig. \ref{fig2}). In the 40--80 keV band this is probably the case as well, but the large number of upper limits prevents confirmation (see Appendix \ref{B}). It is not surprising to find the same behaviour for the 3--5~keV and the 7--15~keV correlations since they are sampling the featureless AGN continuum. We might have expected a different behaviour for the 15--40~keV correlation because it includes the Compton hump, but that is not the case. A detailed spectral analysis of the {\textit{NuSTAR}} data is beyond the scope of this paper, and will be presented elsewhere (Balokovi\'c et al. in preparation). However, that study shows that the contribution of the Compton hump for the galaxies in our sample is in the range 10-20\%. This would be consistent with the result that we see no significant differences in the correlations between the different X-ray energy bands.

We finally studied the strength of the correlation between the X-ray and the MIR emission-line luminosities. All the lines considered show a good correlation with the X-rays ($\rho\ge$0.7), and the correlation strength does not depend on the IP of the lines (see Fig. \ref{fig3}). For example, the strongest correlation is found for the low-IP line [Ne\,II], which is commonly used as a star formation tracer in galaxies. Nevertheless, according to \citet{Melendez08b}, this emission line in active galaxies has both AGN and star formation contributions, and the latter is likely to be insignificant in the AGN-dominated systems that make up our sample. In principle, we expected the lines with the highest IPs ([Ne\,V] and [O\,IV]) to show the strongest correlations with the X-rays, but the correlation coefficients are intermediate between those of lower IPs. As recently shown by \citet{Landt15a,Landt15b} for the type 1 AGN NGC\,4151 and NGC\,5548, both included in our sample, variability in the coronal line region could be a potential source of ``dilution'' of the correlation coefficient strengths, explaining the results found for these high-IP lines.

\section{Conclusions}
\label{Conclusions}
In this work we apply, for the first time, the correlation spectrum technique to cross-correlate 
monochromatic NIR and MIR luminosities with various {\textit{NuSTAR}} X-ray bands for a sample of 24 unobscured type 1 AGN. The main results are as follows:\\

\textbullet \ \ The {\textit{NuSTAR}} data studied here allows to divide the X-ray emission up into selected energy bands between 3 and 80 keV. We find both the shape of the IRXCS and the correlation strengths independent of the X-ray band selected, although the large number of upper limits in the 40--80 keV band prevents confirmation at the highest energies. The same situation is apparent when considering the emission line correlations.

\textbullet \ \ We find the IRXCS correlation maximum at $\sim$15-20~$\mu$m. This peak coincides with the maximum contribution of the AGN to the IRS spectra analyzed here. 
At $\lambda\gtrsim$20~$\mu$m the correlation spectrum slowly decreases.

\textbullet \ \ We find a NIR correlation peak at $\sim$2~$\mu$m, which we associate with the NIR bump observed in the SEDs of some type 1 AGN at $\sim$1--5~$\mu$m. The results presented here indicate that this correlation feature is produced by AGN-heated dust. 

\textbullet \ \ All the MIR emission lines considered here, whose IPs range from 8 to 97~eV, show a good correlation with the hard X-rays ($\rho\ge$0.7). It is noteworthy that the correlation strength does not depend on the IP of the lines.\\

Detailed studies such as this, carried out across the whole NIR-MIR spectral range, are clearly complementary to very large statistical studies, for example \citet{Chen17},  which focus on specific X-ray and IR bands, but with wider luminosity and redshift coverage. In the future the spectral correlation technique may be applied to AGN samples observed with the combined spectral coverage of NIRSpec and MIRI aboard the {\textit{James Webb Space Telescope (JWST)}}.

\section*{Acknowledgments}

IGB acknowledges financial support from the Instituto de Astrof\'isica de Canarias through Fundaci\'on La Caixa. IGB also acknowledges the Durham University for their hospitality during his stay in March and April 2015 when this project was started. CRA acknowledges the Ram\'on y Cajal Program of the Spanish Ministry of Economy and Competitiveness through project RYC-2014-15779. M.\,B. acknowledges support from NASA Headquarters under the NASA Earth and Space Science Fellowship Program, grant NNX14AQ07H. We finally acknowledge useful comments from the anonymous referee.

\appendix
\section{Spectral Decomposition}
\label{A}

Here we show the {\textit{Spitzer/IRS}} spectra of our sample of 24 unobscured type 1 AGN, which were retrieve form the CASSIS atlas \citep{Lebouteiller11}. Considering the spatial scales probed (0.2\,kpc~$\lesssim$ spatial scale $\lesssim$~5.4\,kpc), we expect some degree of contribution from the host galaxy to the MIR spectra, in addition to the AGN. To estimate these contributions, we use the DeblendIRS routine \citep{Hernan-caballero2015}, that decomposes MIR spectra using a linear combination of three spectral components: AGN, PAH and stellar emission. The spectrum of each component is selected from a large library containing {\textit{Spitzer/IRS}} spectra from extreme cases of galaxies dominated by AGN, PAH, or stellar emission. Note that we have consistently avoided the use of the sources themselves as templates, since few of our galaxies are in the AGN library of DeblendIRS. Due to limitations in the library, we cannot expand the spectral decomposition beyond 22~$\mu$m. A detailed description of the method is given in \citet{Hernan-caballero2015}, which we already tested in \citet{Bernete16}.

We present the main results and properties derived from the spectral decomposition of our sample in Fig. \ref{fig5} and Table \ref{tab_appendix}. We found that the median value of the factional contribution of the AGN to the {\textit{Spitzer/IRS}} spectra is 0.90, as expected for type 1 AGN. For the majority of the galaxies we find the maximum of the fitted AGN template at 20-25~$\mu$m. Note that in the case of NGC\,7213 we cannot obtain a reliable fit, as the modelling does not reproduce the 9.7~$\mu$m silicate emission feature.
\begin{table*}
\centering
\begin{tabular}{lcccccccccc}
\hline
Name 		&	F$_{AGN}$&	F$_{PAH}$&	F$_{Stellar}$&	$\alpha_{AGN}$&	S$_{Si}$ &	F$_{6\mu m}^{\,AGN}$ & F$_{12\mu m}^{\,AGN}$ & F$_{20\mu m}^{\,AGN}$ \\
\hline
Mrk\,335	&0.82	&0.04	&0.14	&-1.30	&0.25	&0.64	&0.86	&0.92	\\
Fairall\,9	&0.91	&0.02	&0.07	&-1.22	&0.19	&0.82	&0.94	&0.97	\\
Mrk\,1018	&0.83	&0.02	&0.15	&-1.16	&0.24	&0.67	&0.88	&0.90	\\
Mrk\,590	&0.88	&0.05	&0.07	&-2.74	&0.32	&0.59	&0.92	&0.98	\\
Mrk\,1044	&0.85	&0.06	&0.09	&-1.22	&0.19	&0.75	&0.89	&0.93	\\
3C\,120		&0.96	&0.04	&0.00	&-1.62	&0.18	&0.97	&0.97 	&0.95 	\\
Ark\,120	&0.95	&0.05	&0.00	&-0.64	&0.28	&0.98	&0.95 	&0.88	\\
1H\,0707-495 &0.94	&0.03	&0.03	&-1.30	&0.25	&0.89	&0.95	&0.94	\\
RBS\,0770	&0.99	&0.01	&0.00	&-1.24	&0.12	&0.99	&1.00	&0.99	\\
NGC\,4051	&0.91	&0.07	&0.02	&-1.75	&0.07	&0.83	&0.92	&0.97	\\
NGC\,4151	&0.93	&0.02	&0.05	&-1.86	&0.03	&0.81	&0.95	&0.96	\\
PG\,1211+143 &0.97	&0.02	&0.01	&-1.22	&0.19	&0.96	&0.98	&0.96	\\
NGC\,4593	&0.84	&0.01	&0.15	&-1.19	&-0.03	&0.65	&0.89	&0.92	\\
Mrk\,766	&0.83	&0.07	&0.10	&-2.27	&-0.17	&0.57	&0.88	&0.95	\\
MCG-06-30-015 &0.82	&0.03	&0.15	&-1.91	&-0.19	&0.54	&0.88 	&0.92	\\
IC\,4329A	&0.77	&0.06	&0.17	&-1.86	&0.03	&0.49	&0.82	&0.93	\\
CGCG\,017-073 &0.86	&0.05	&0.09	&-1.75	&0.29	&0.69	&0.90	&0.94	\\
NGC\,5548	&0.92	&0.06	&0.02	&-1.86	&0.06	&0.88	&0.93	&0.93	\\
Mrk\,1393	&0.77	&0.03	&0.20	&-2.67	&-0.09	&0.35	&0.84 	&0.91	\\
Mrk\,290	&0.93	&0.01	&0.06	&-1.66	&0.17	&0.80	&0.96	&0.97	\\
3C\,382		&0.77	&0.10	&0.13	&-0.30	&0.33	&0.74	&0.79	&0.63	\\
3C\,390.3	&0.90	&0.02	&0.08	&-1.66	&0.17	&0.73	&0.93	&0.96	\\
NGC\,7213	&0.96	&0.00	&0.04	&-1.92	&0.33	&0.90	&0.97 	&0.98	\\
Mrk\,915	&0.84	&0.04	&0.12	&-2.79	&0.20	&0.53	&0.88	&0.93	\\
\hline
\end{tabular}						 
\caption{Properties derived from the spectral decomposition of the {\textit{Spitzer/IRS}} spectra of the sample. Columns 2, 3 and 4 correspond to the fractional contribution of the AGN, the PAH and the stellar components to the MIR spectrum. Columns 5 and 6 list the spectral index of the AGN component and the silicate strength (positive and negative values correspond to emission and absorption features, respectively). Finally, columns 7 and 9 correspond to the fractional contribution of the AGN component to the rest-frame spectrum at 6~$\mu$m, 12~$\mu$m and 20~$\mu$m, respectively.}
\label{tab_appendix}
\end{table*}

\begin{figure*}
\centering
\par{
\includegraphics[width=8.0cm]{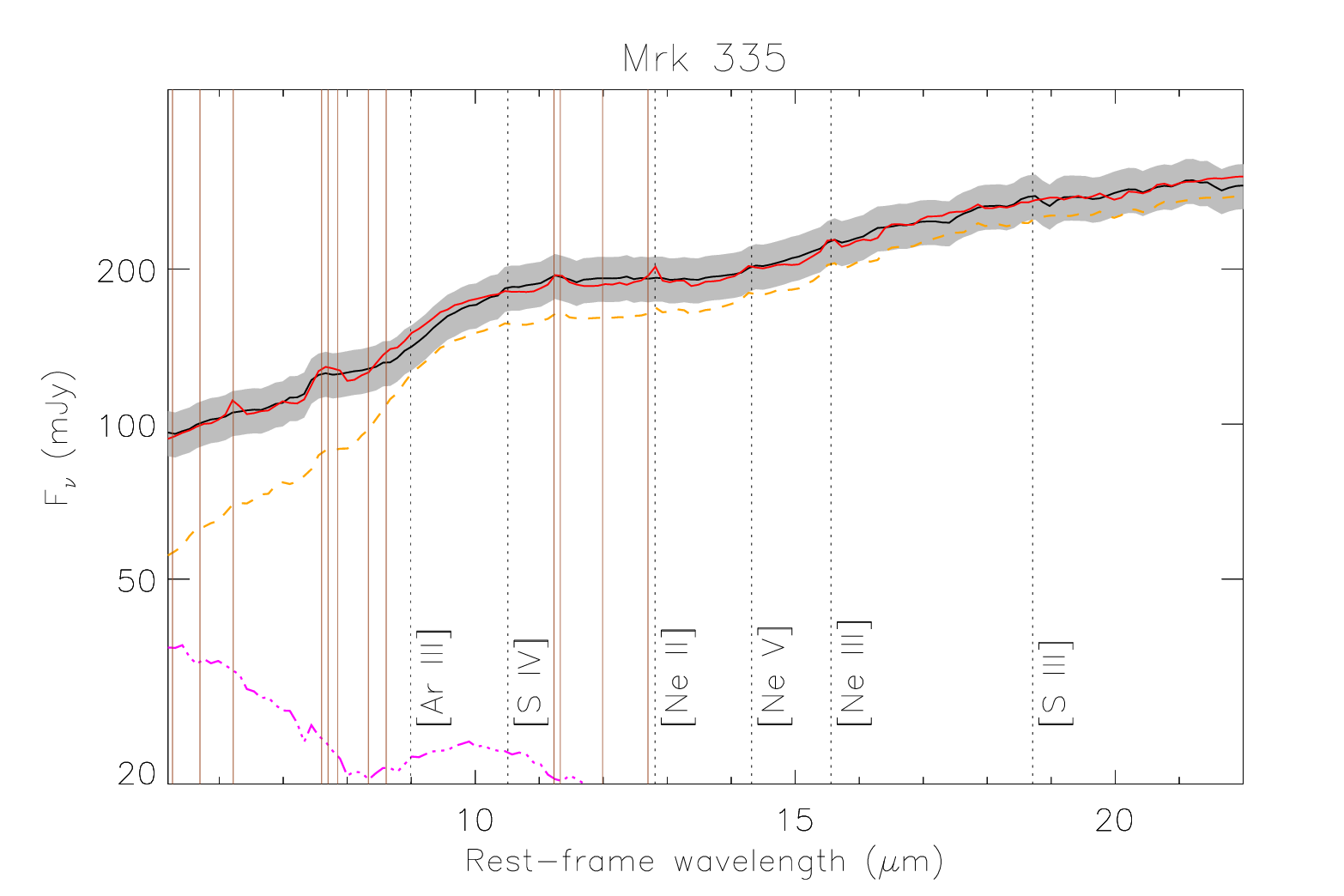}
\includegraphics[width=8.0cm]{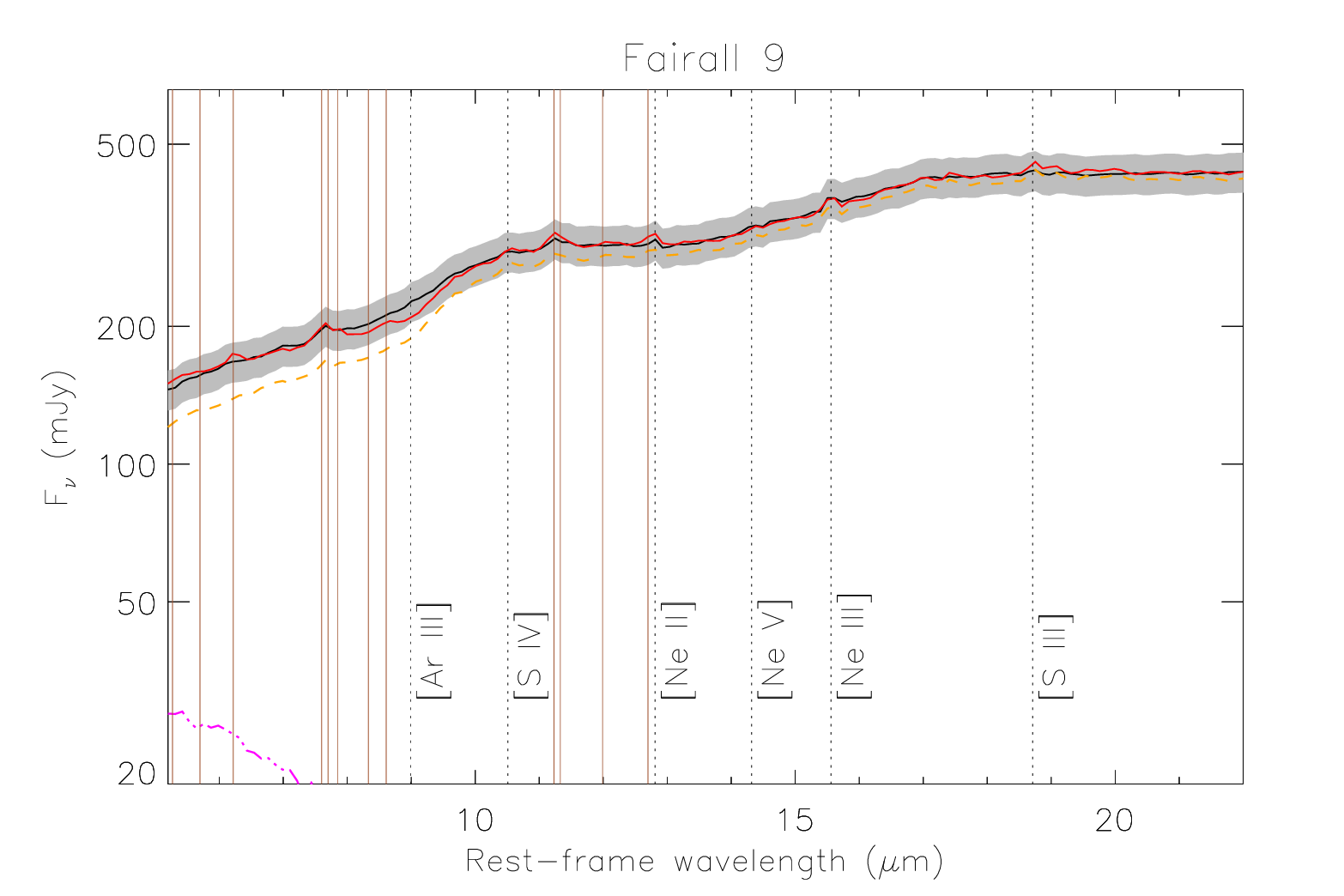}
\includegraphics[width=8.0cm]{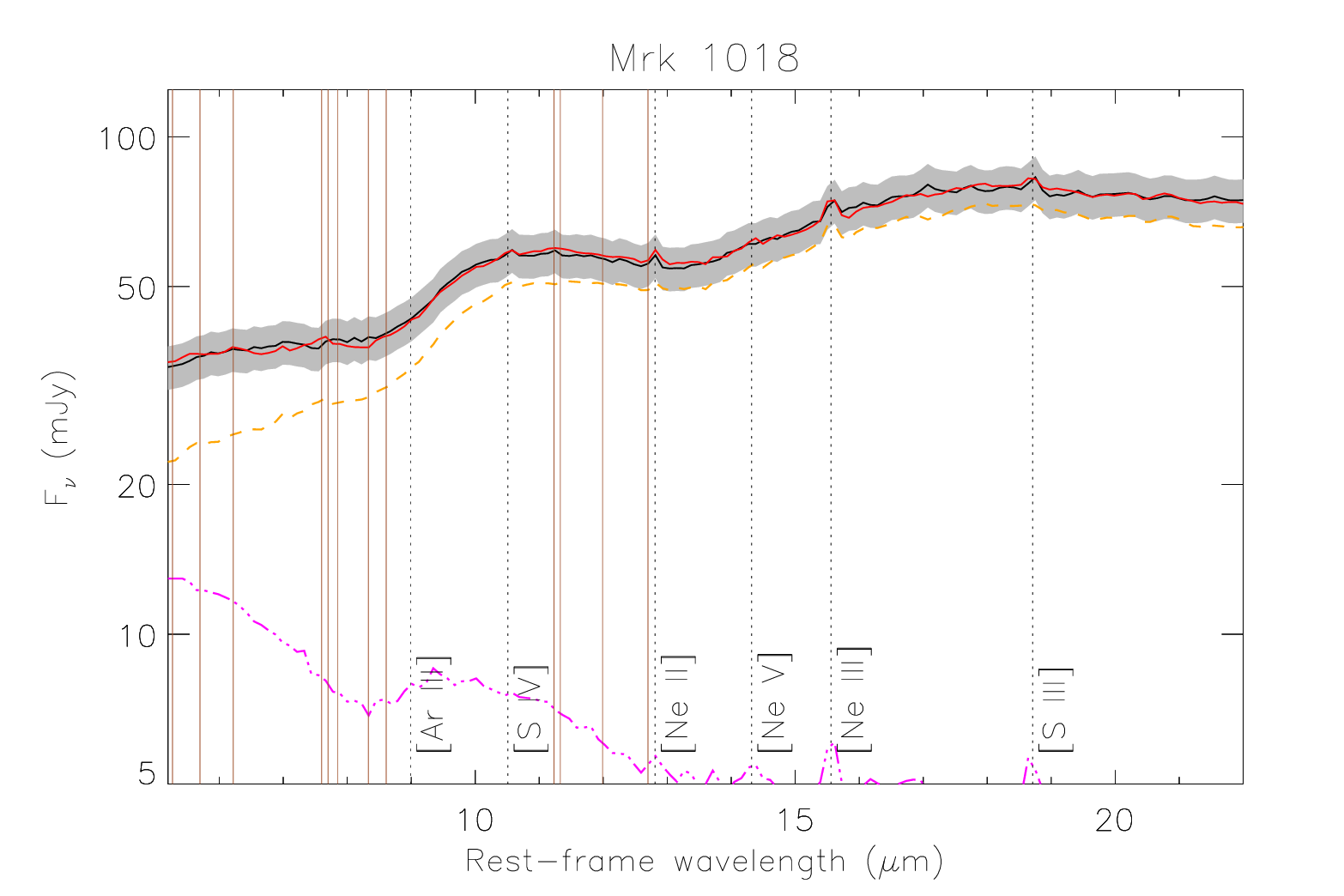}
\includegraphics[width=8.0cm]{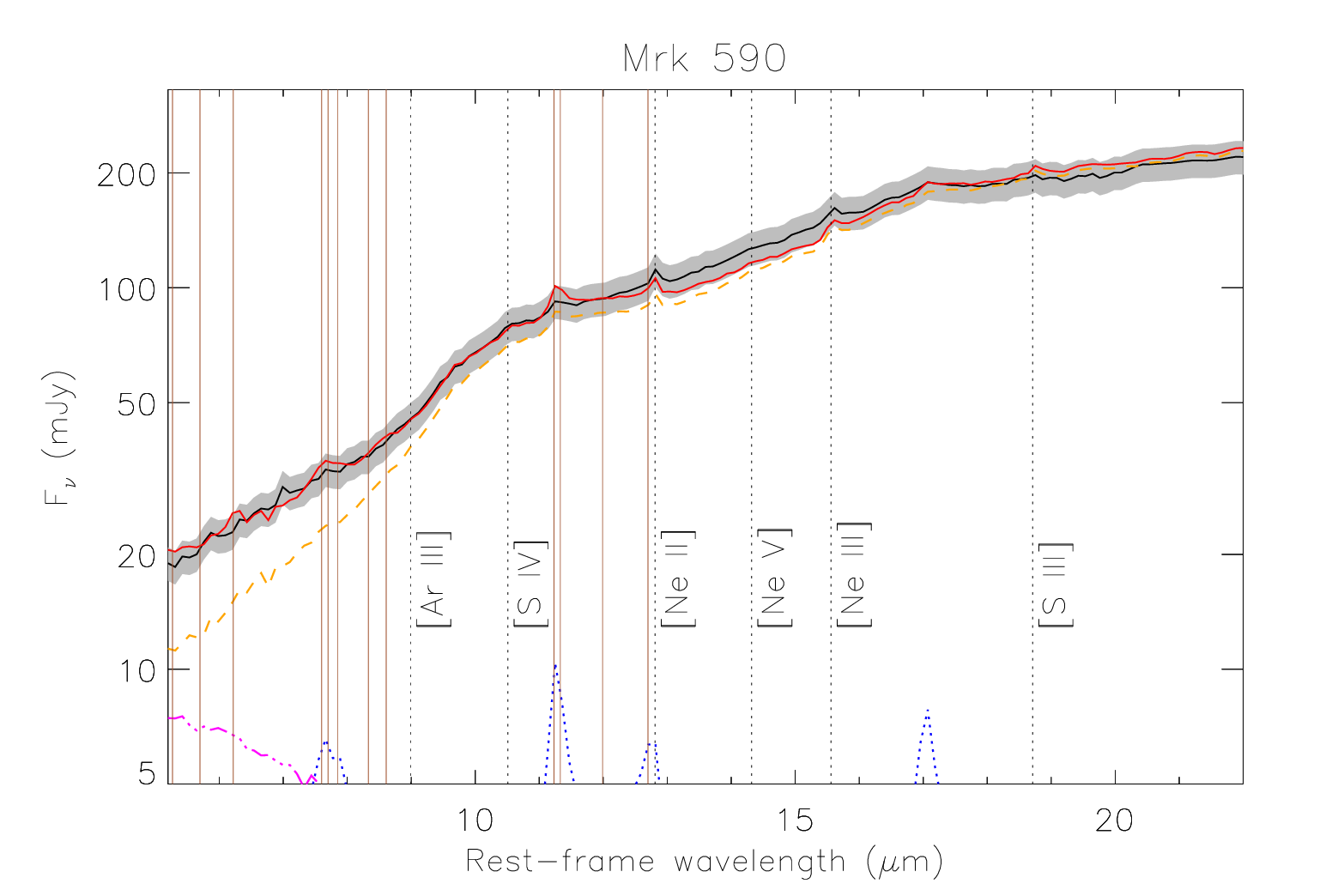}
\includegraphics[width=8.0cm]{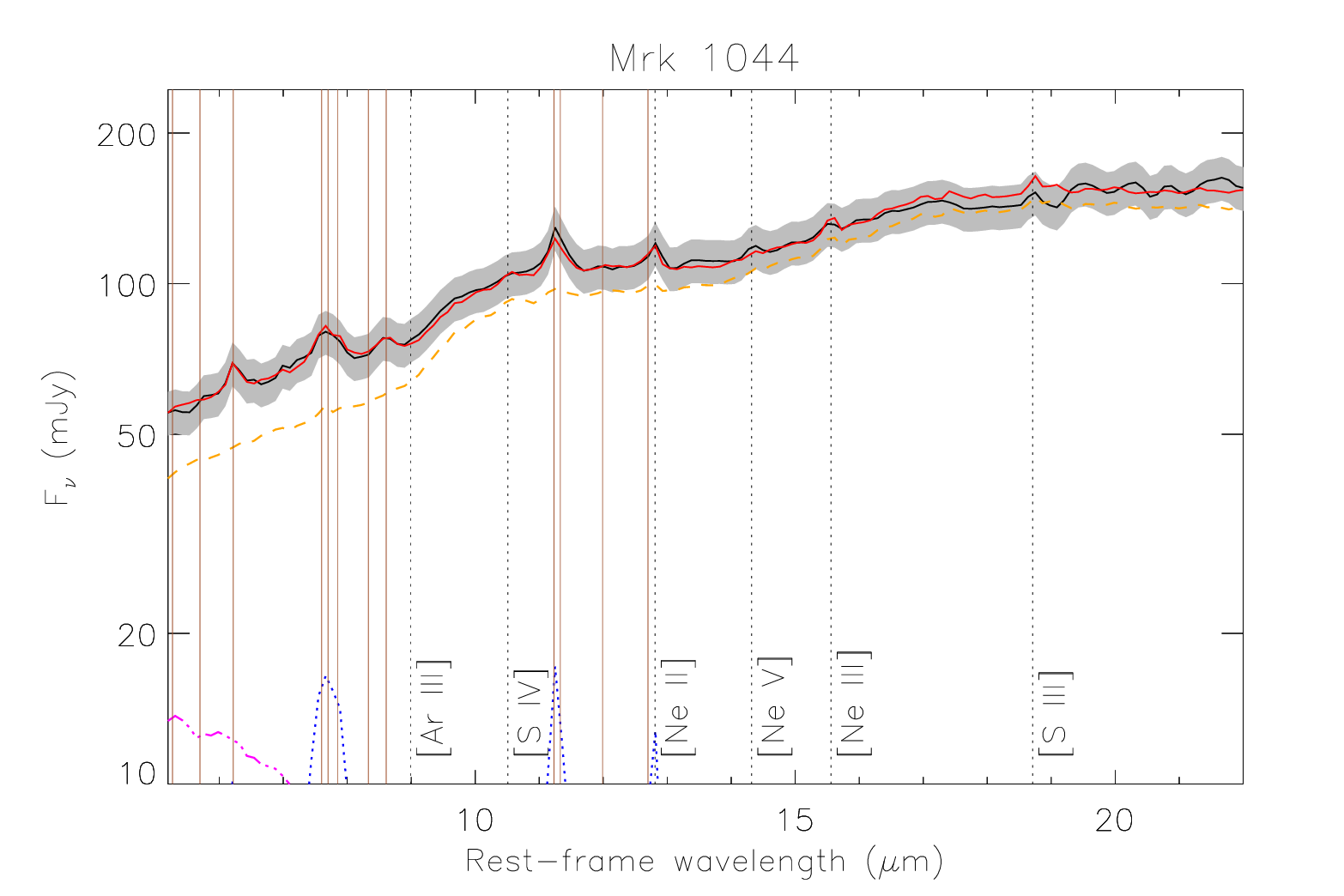}
\includegraphics[width=8.0cm]{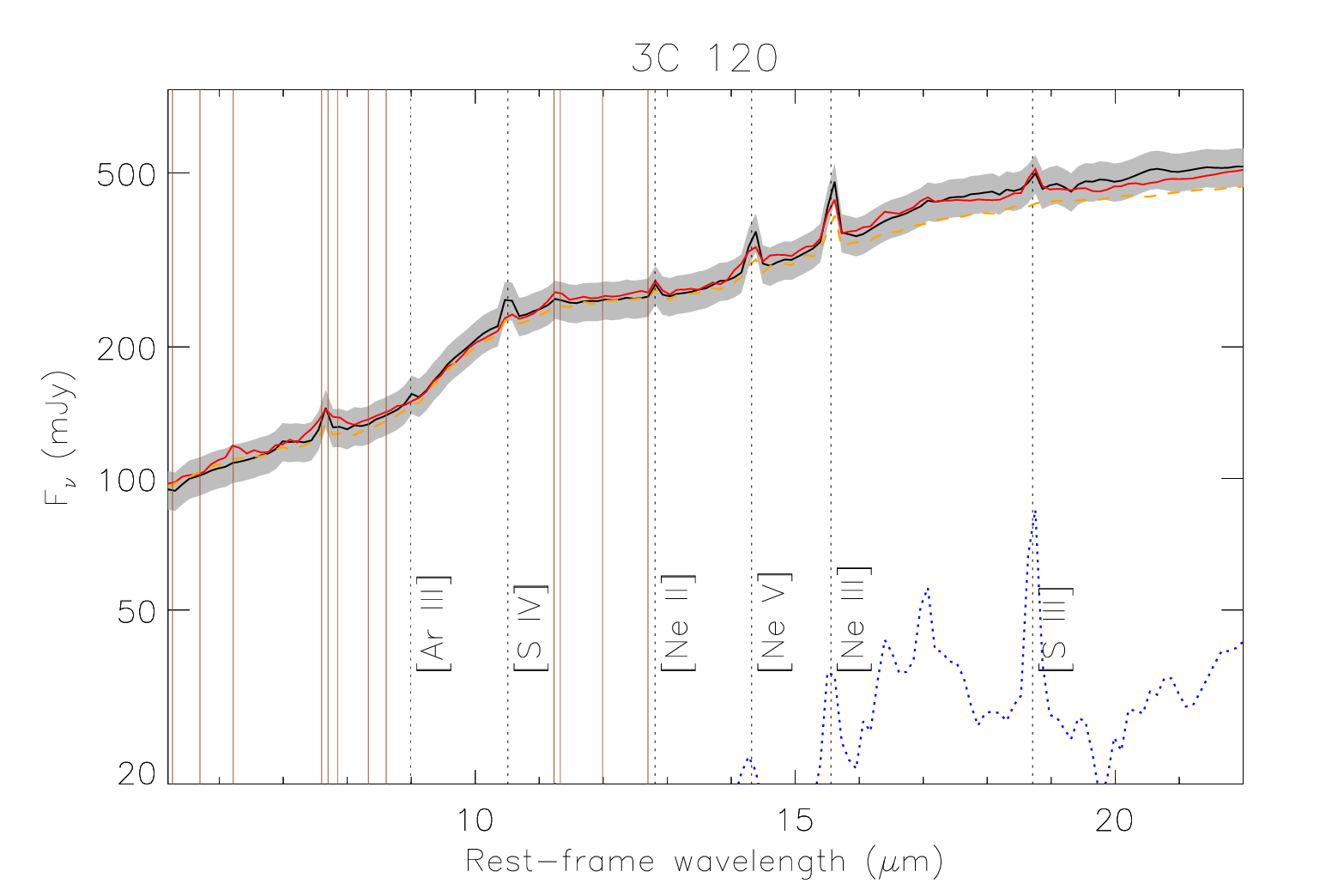}
\includegraphics[width=8.0cm]{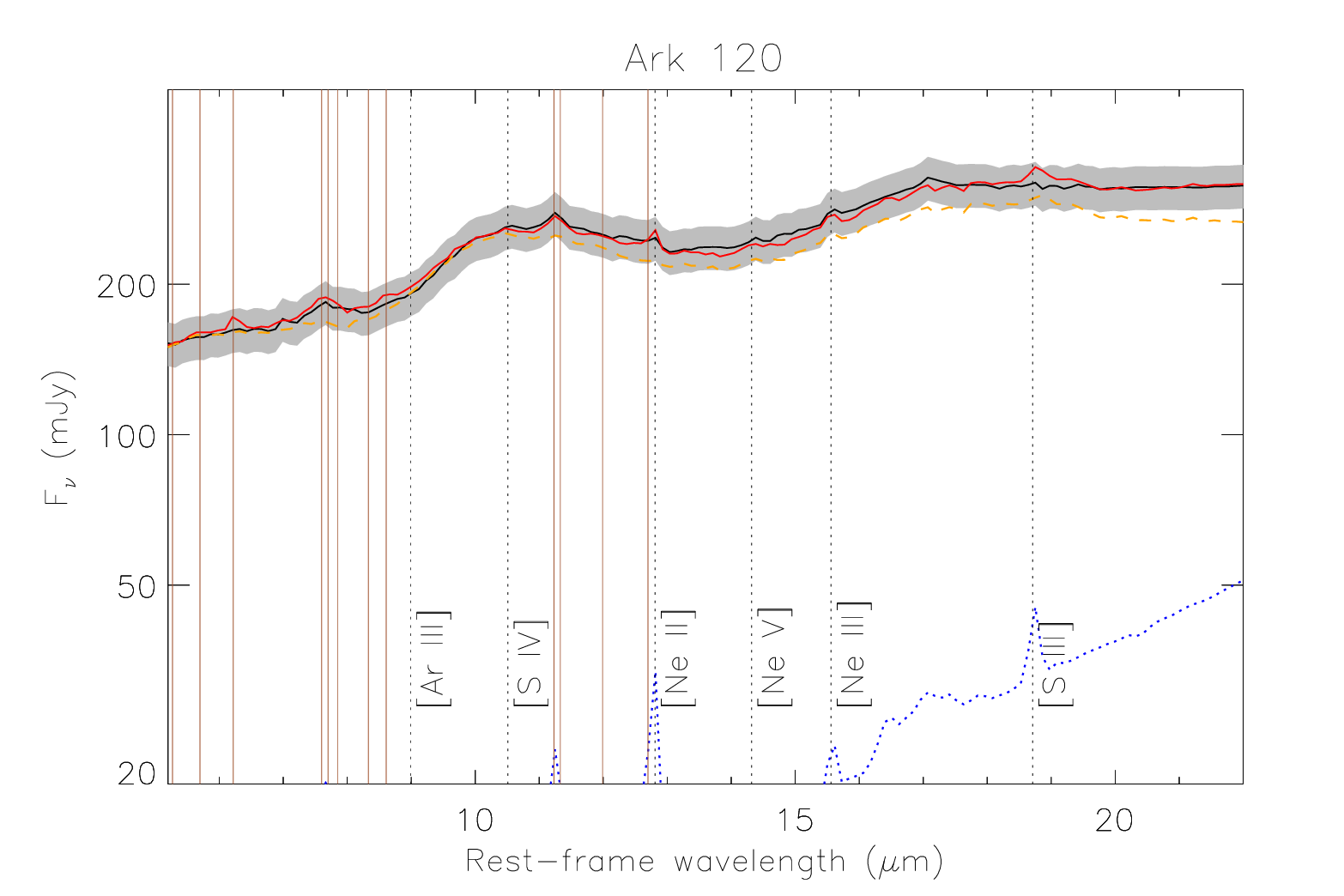}
\includegraphics[width=8.0cm]{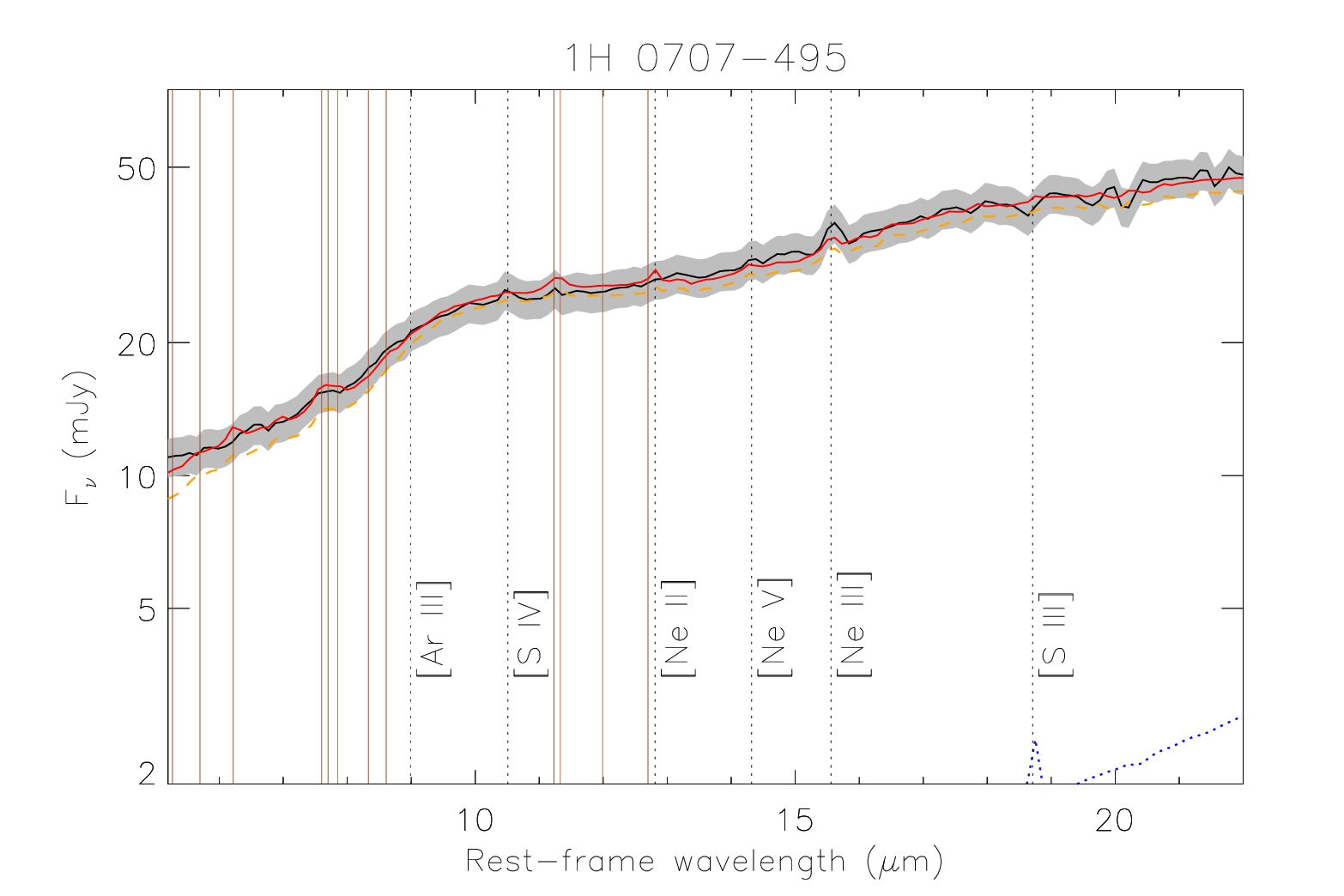}
\par} 
\caption{Spectral decomposition of the {\textit{Spitzer/IRS}} spectra of our sample. We show the rest-frame spectra (black solid lines), best fits (red solid lines), AGN component (orange dashed lines), PAH component (blue dotted lines) and stellar component (dot-dashed fuchsia lines). The brown vertical solid lines correspond to the most important PAH features and the black vertical dotted lines are the main MIR emission lines. The uncertainties of the Spitzer/IRS spectra are shown as grey shaded regions.}
\label{fig5}
\end{figure*}

\begin{figure*}
\contcaption
\centering
\par{
\includegraphics[width=8.0cm]{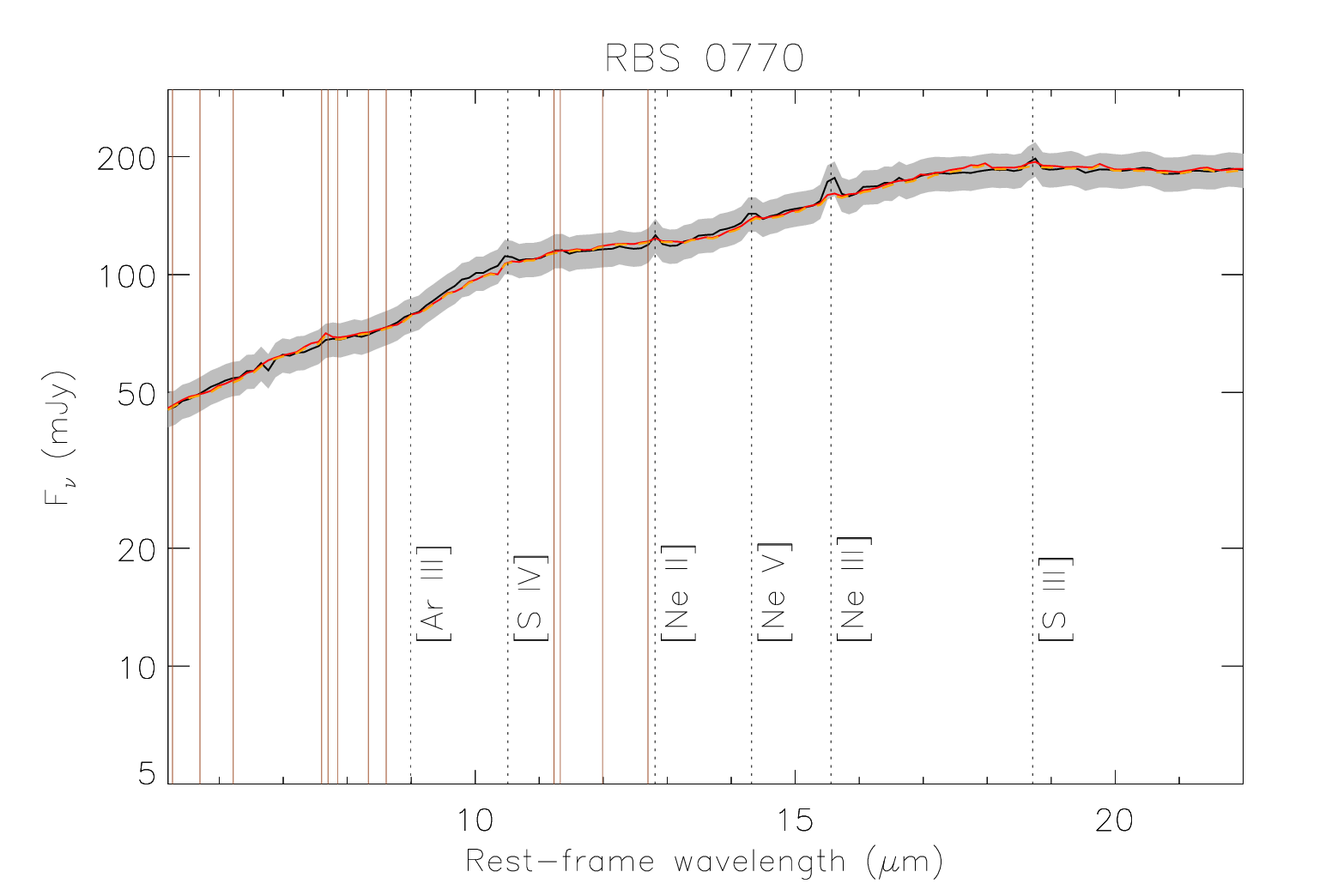}
\includegraphics[width=8.0cm]{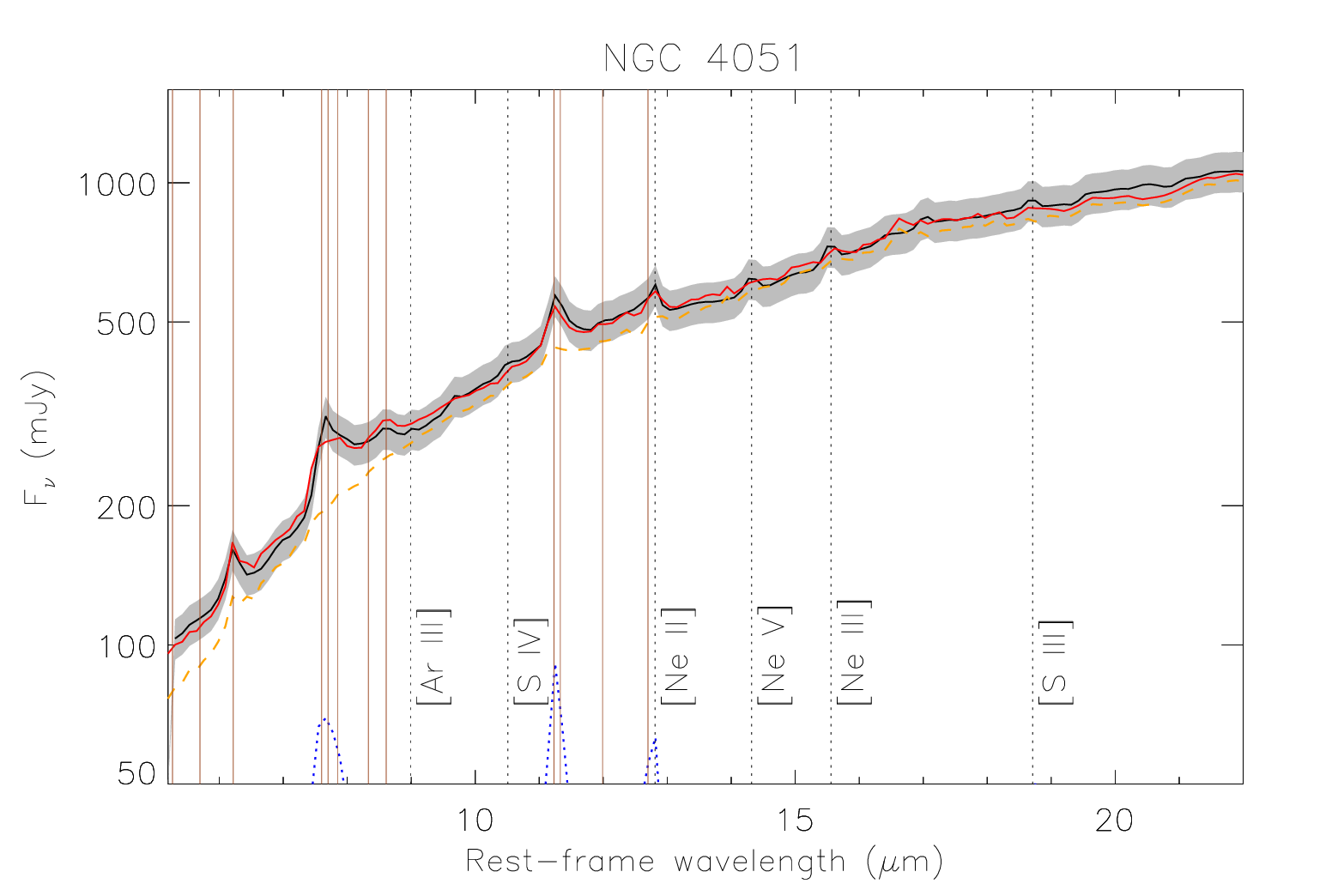}
\includegraphics[width=8.0cm]{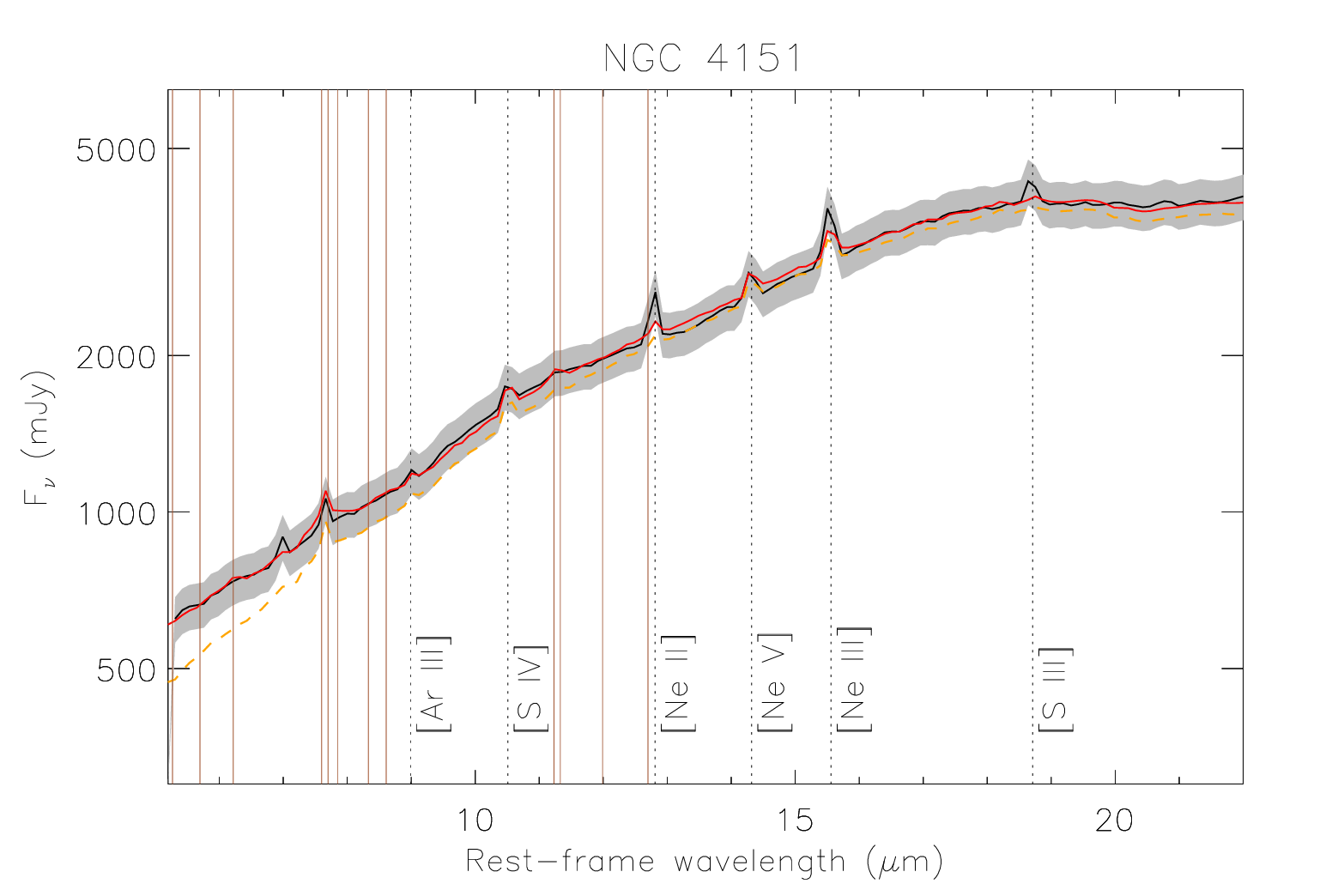}
\includegraphics[width=8.0cm]{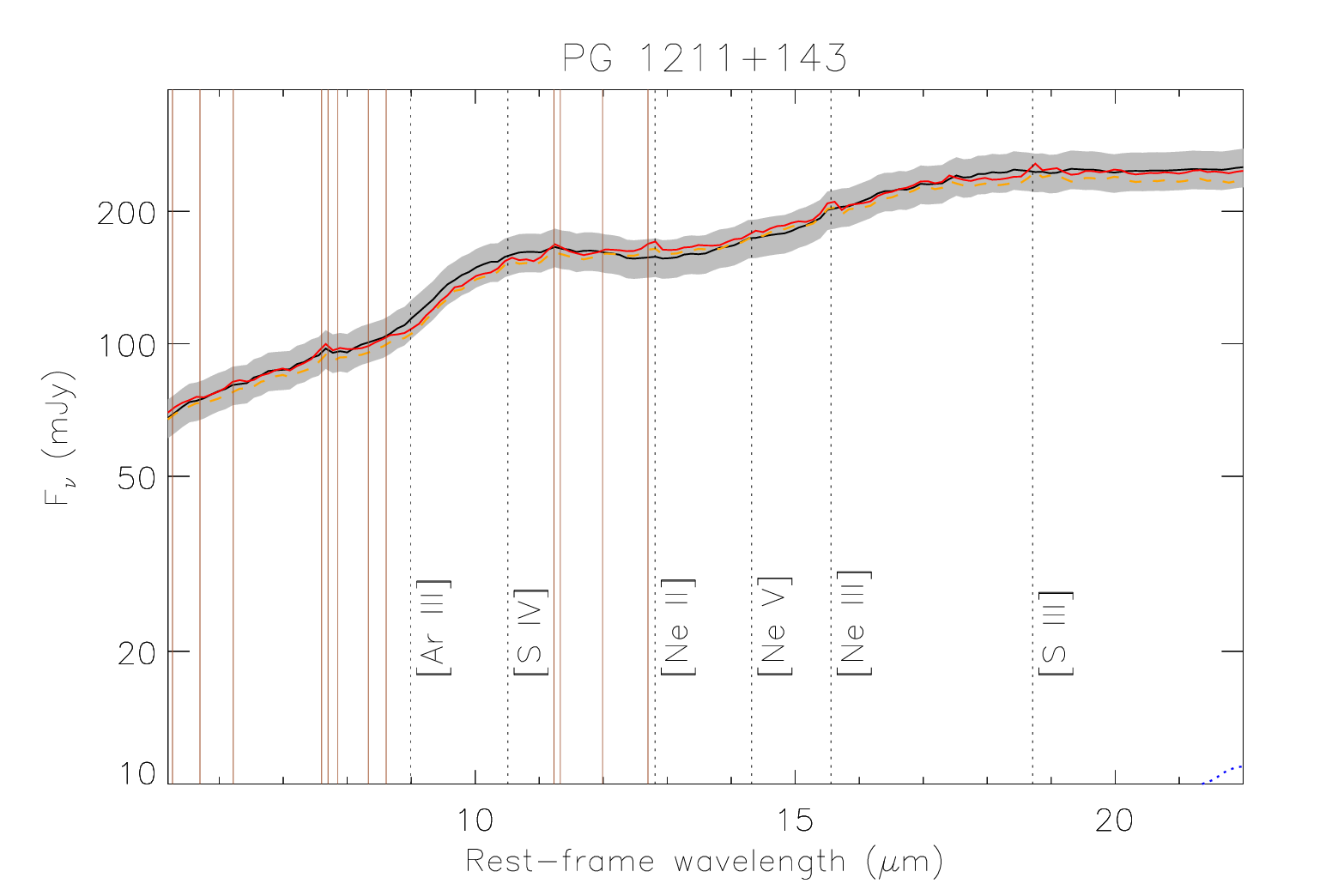}
\includegraphics[width=8.0cm]{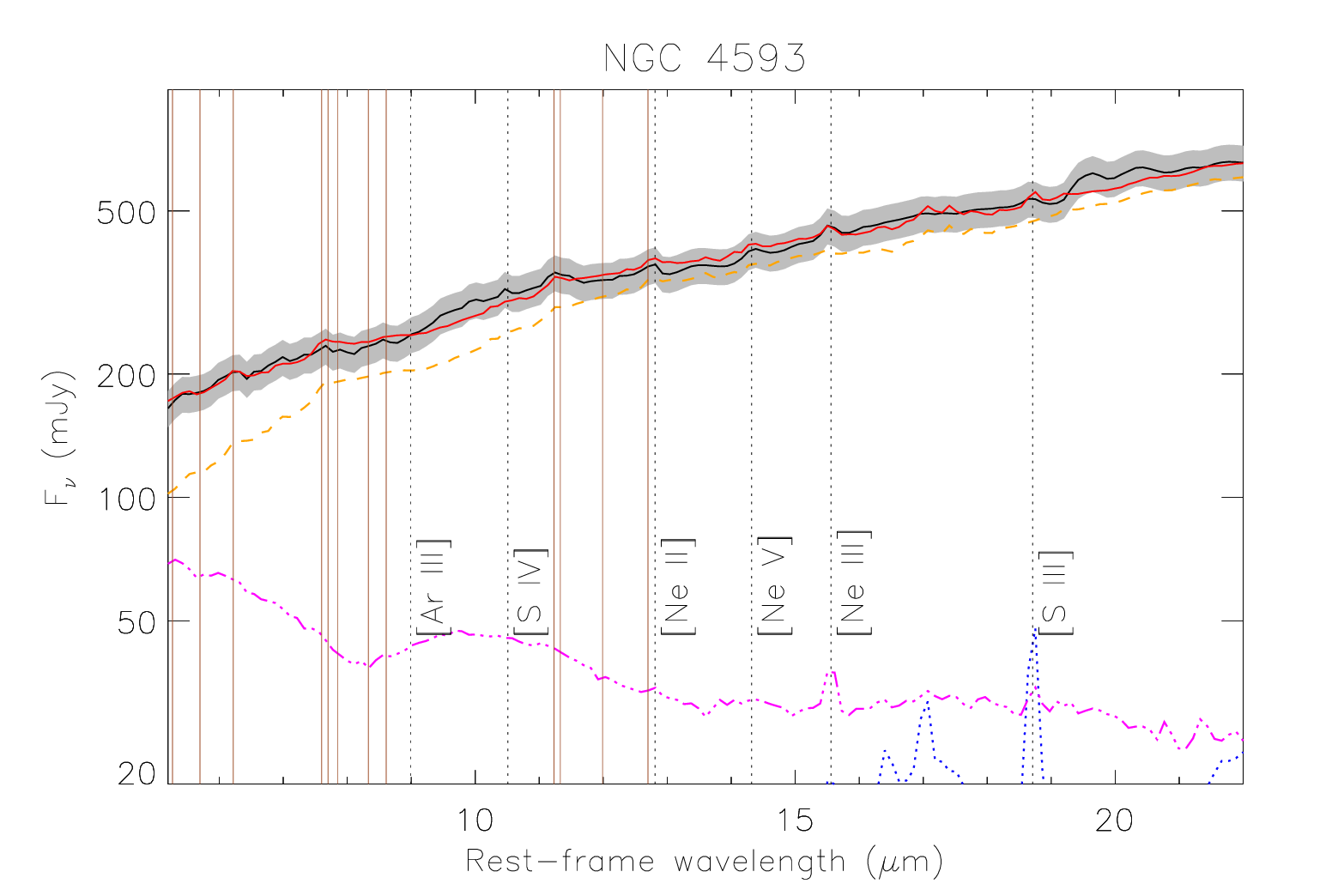}
\includegraphics[width=8.0cm]{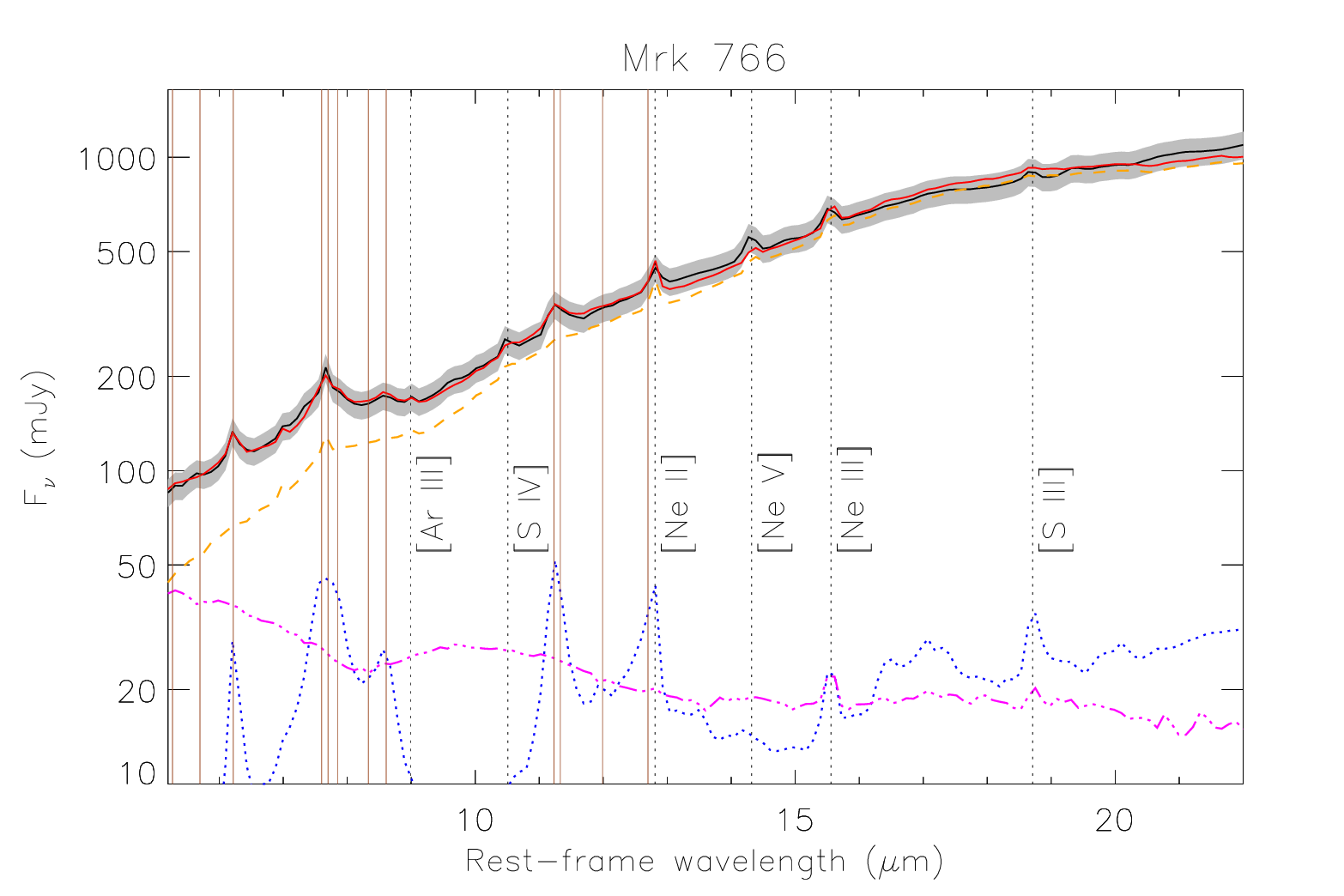}
\includegraphics[width=8.0cm]{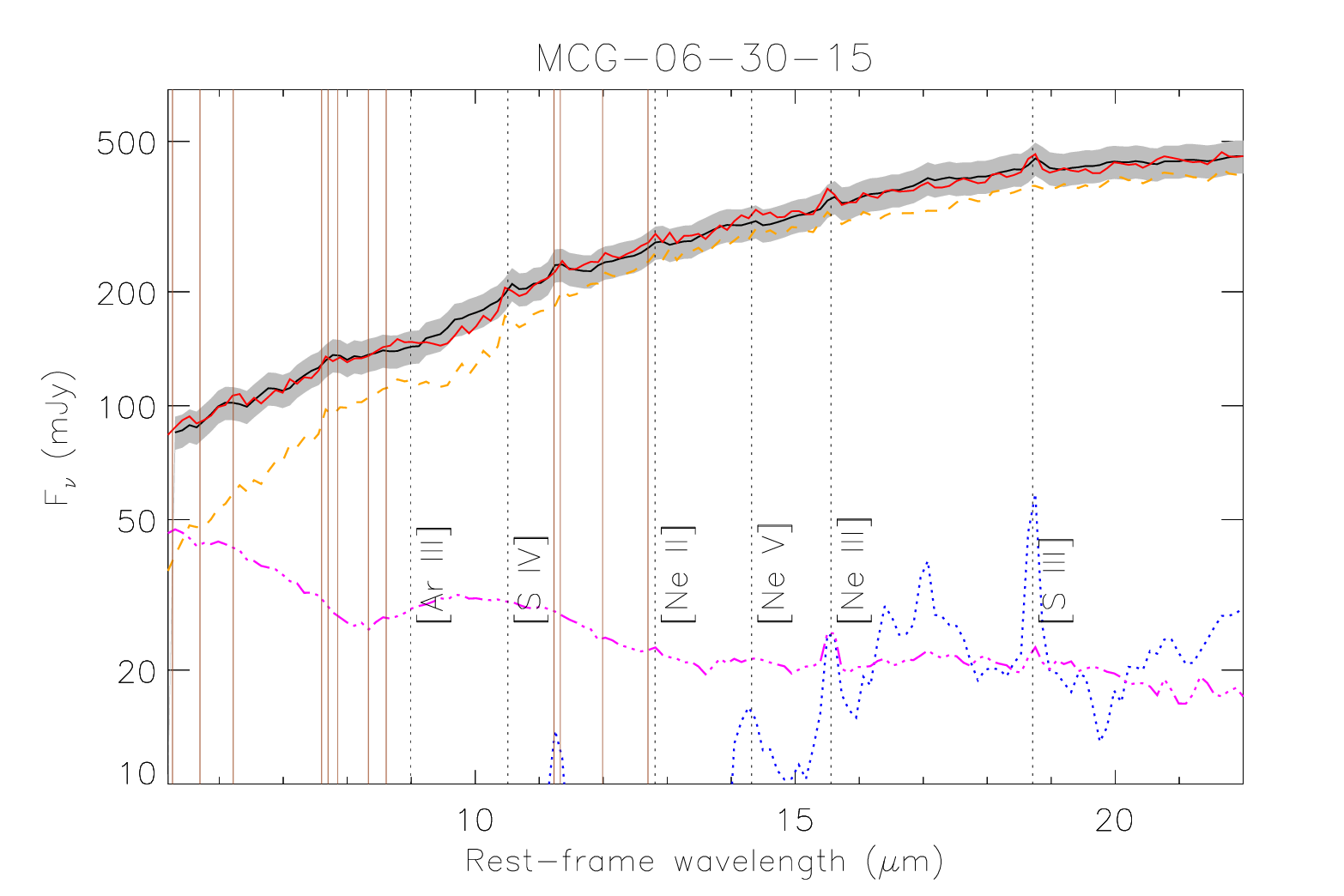}
\includegraphics[width=8.0cm]{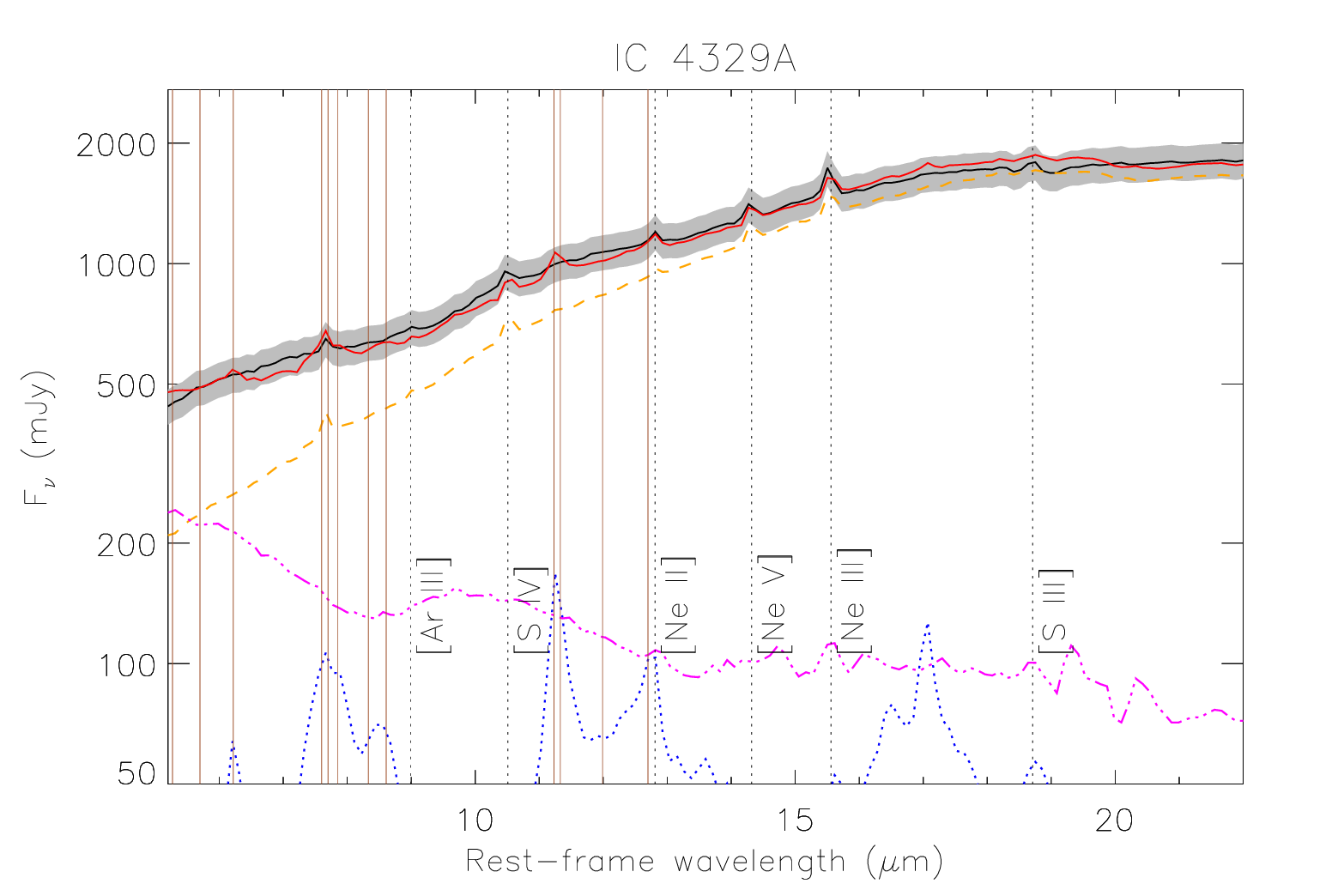}
\par} 
\end{figure*}

\begin{figure*}
\contcaption
\centering
\par{
\includegraphics[width=8.0cm]{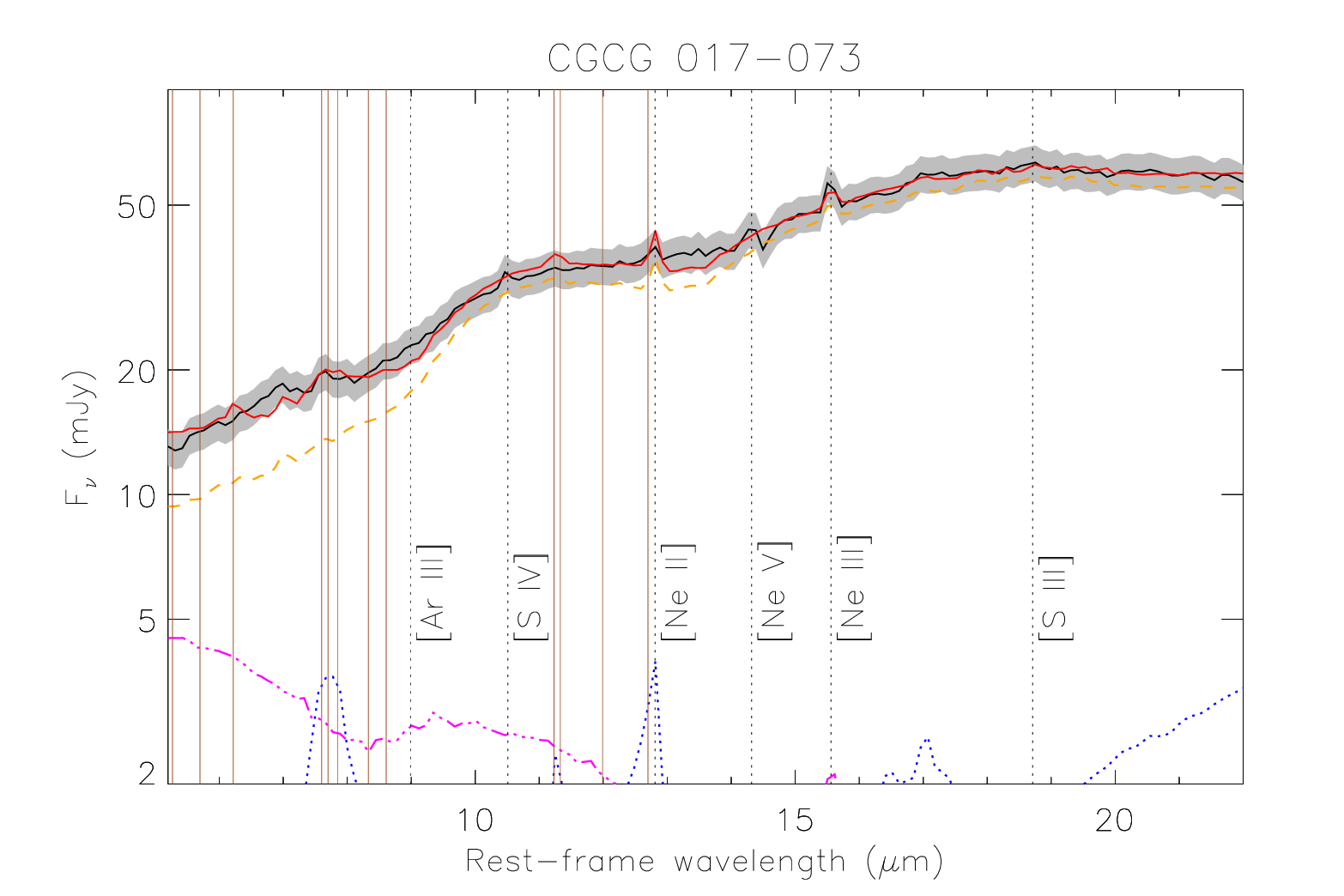}
\includegraphics[width=8.0cm]{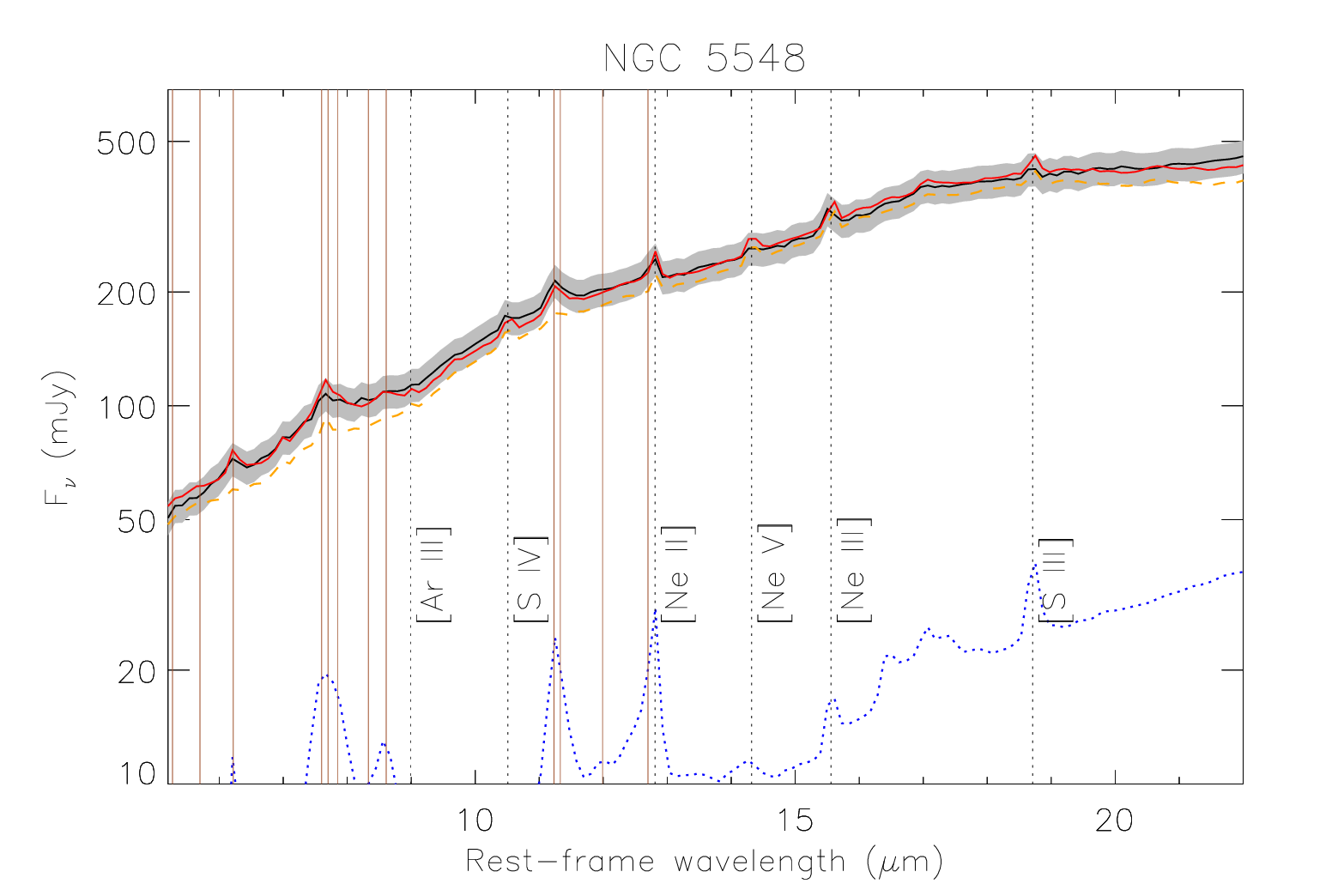}
\includegraphics[width=8.0cm]{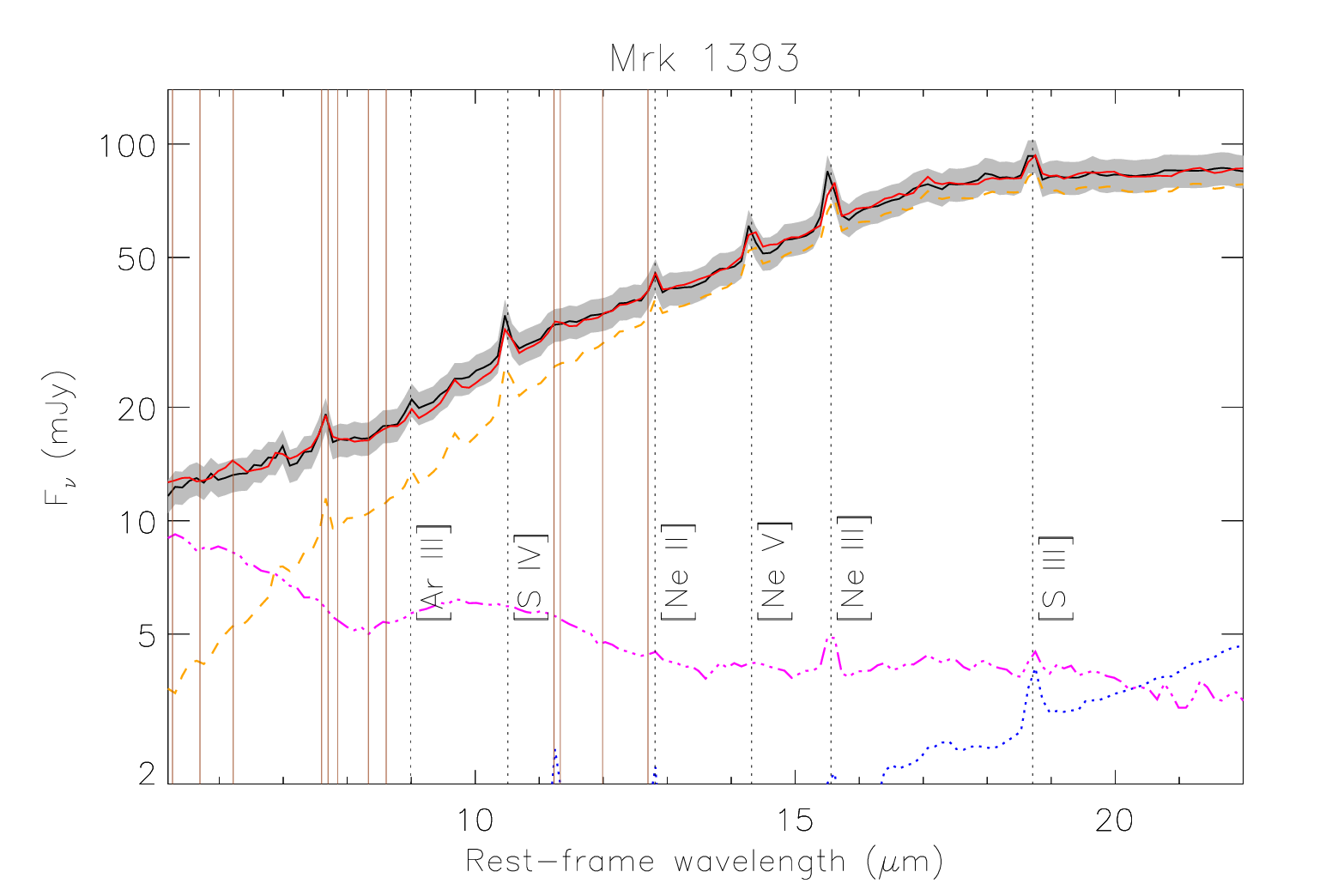}
\includegraphics[width=8.0cm]{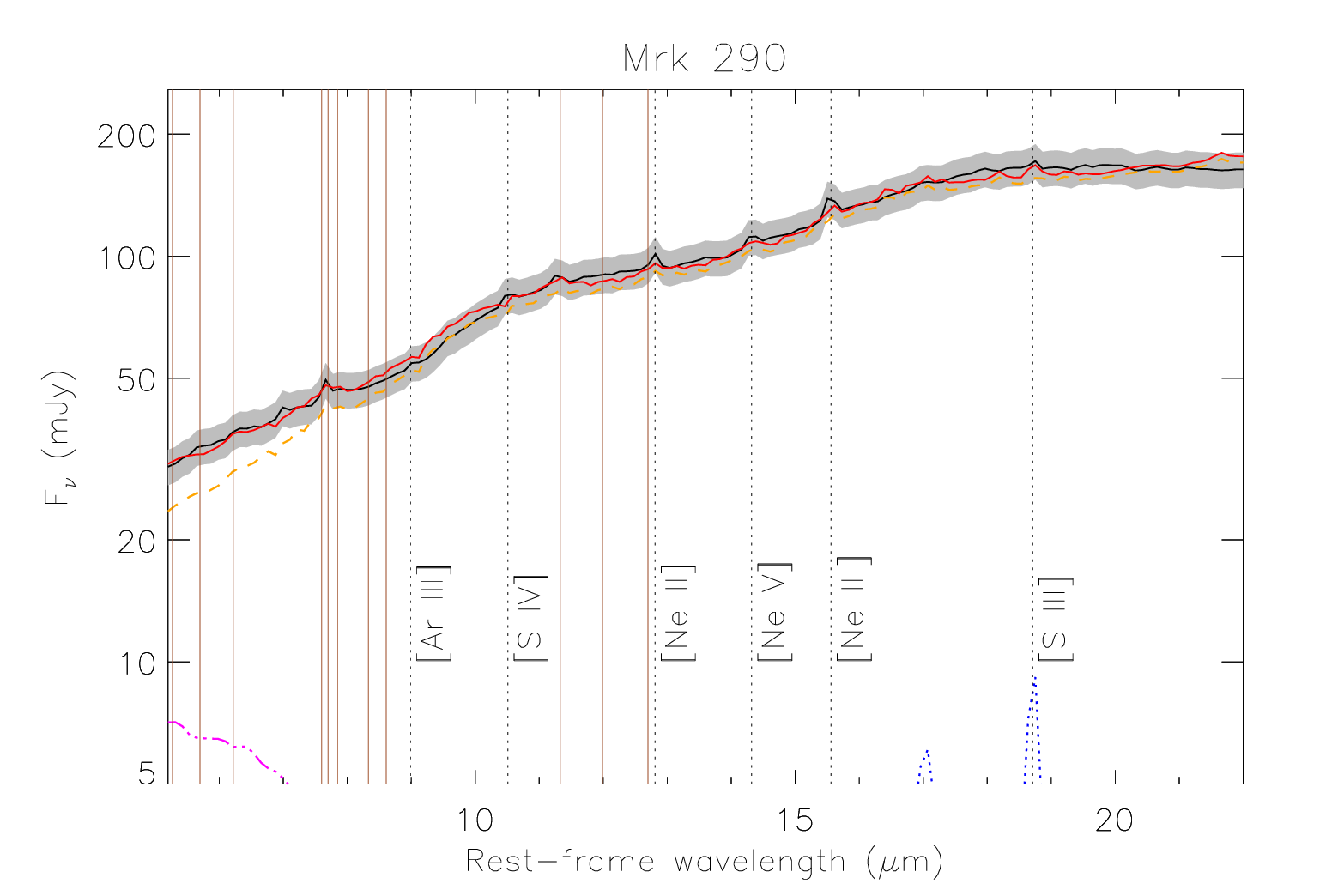}
\includegraphics[width=8.0cm]{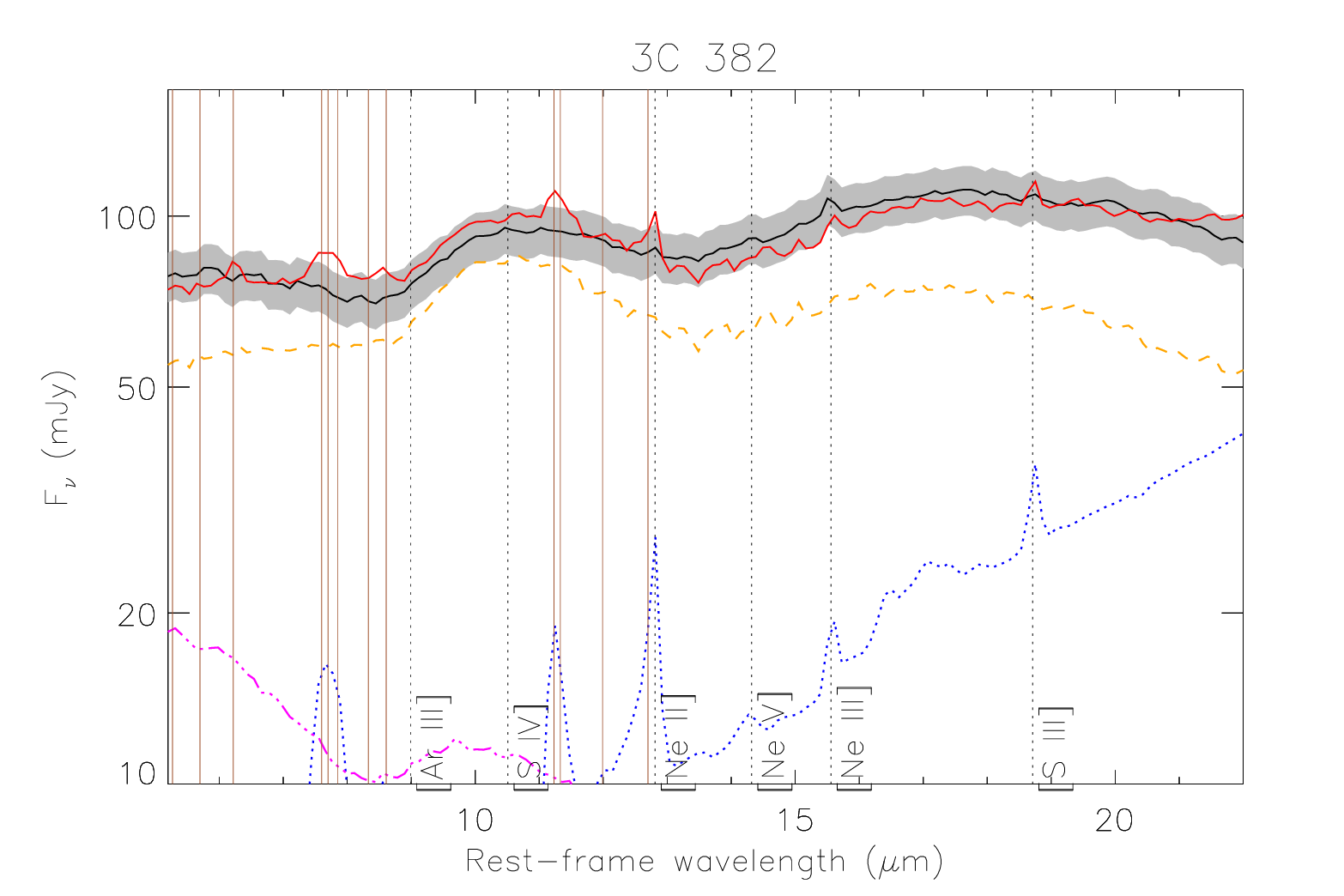}
\includegraphics[width=8.0cm]{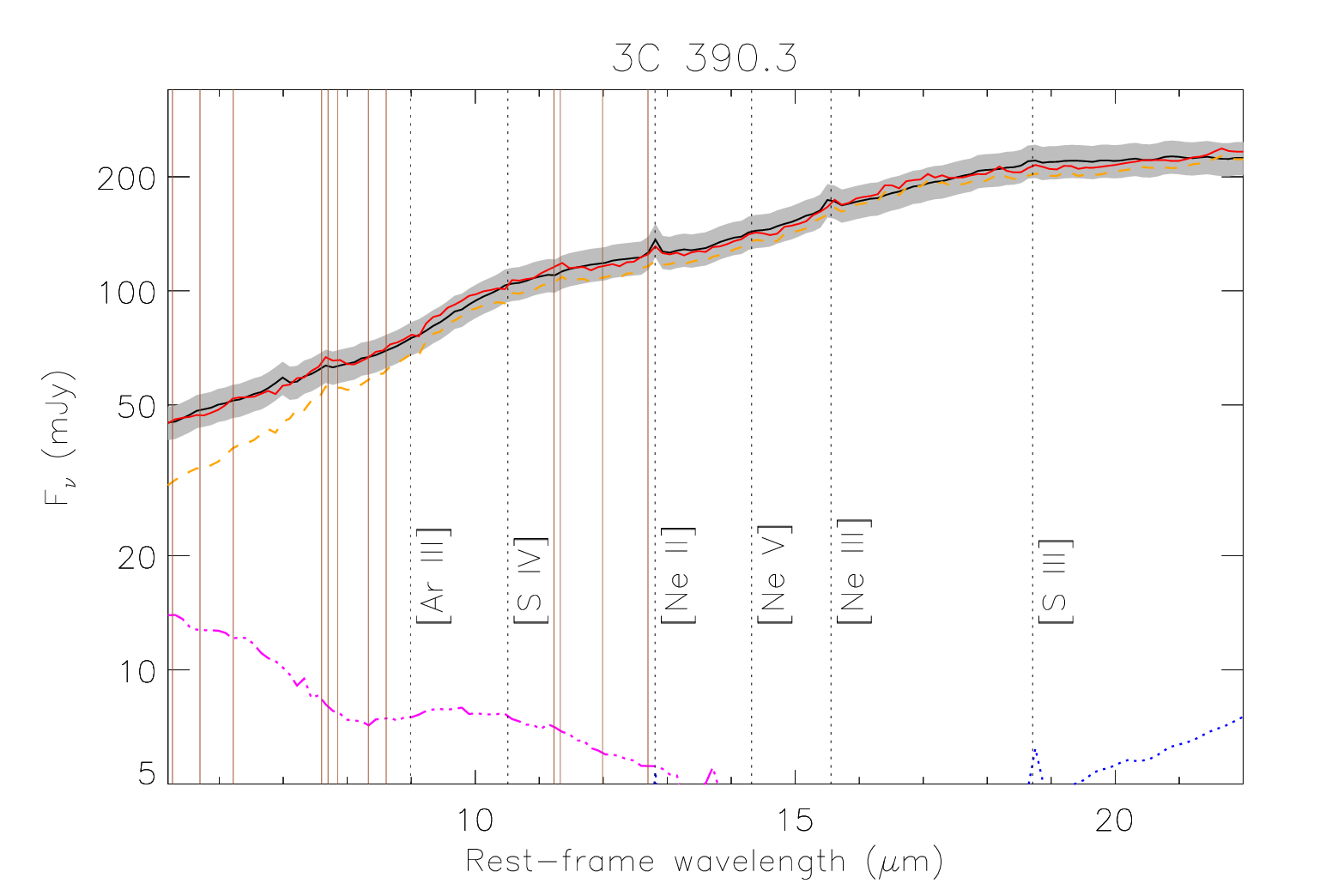}
\includegraphics[width=8.0cm]{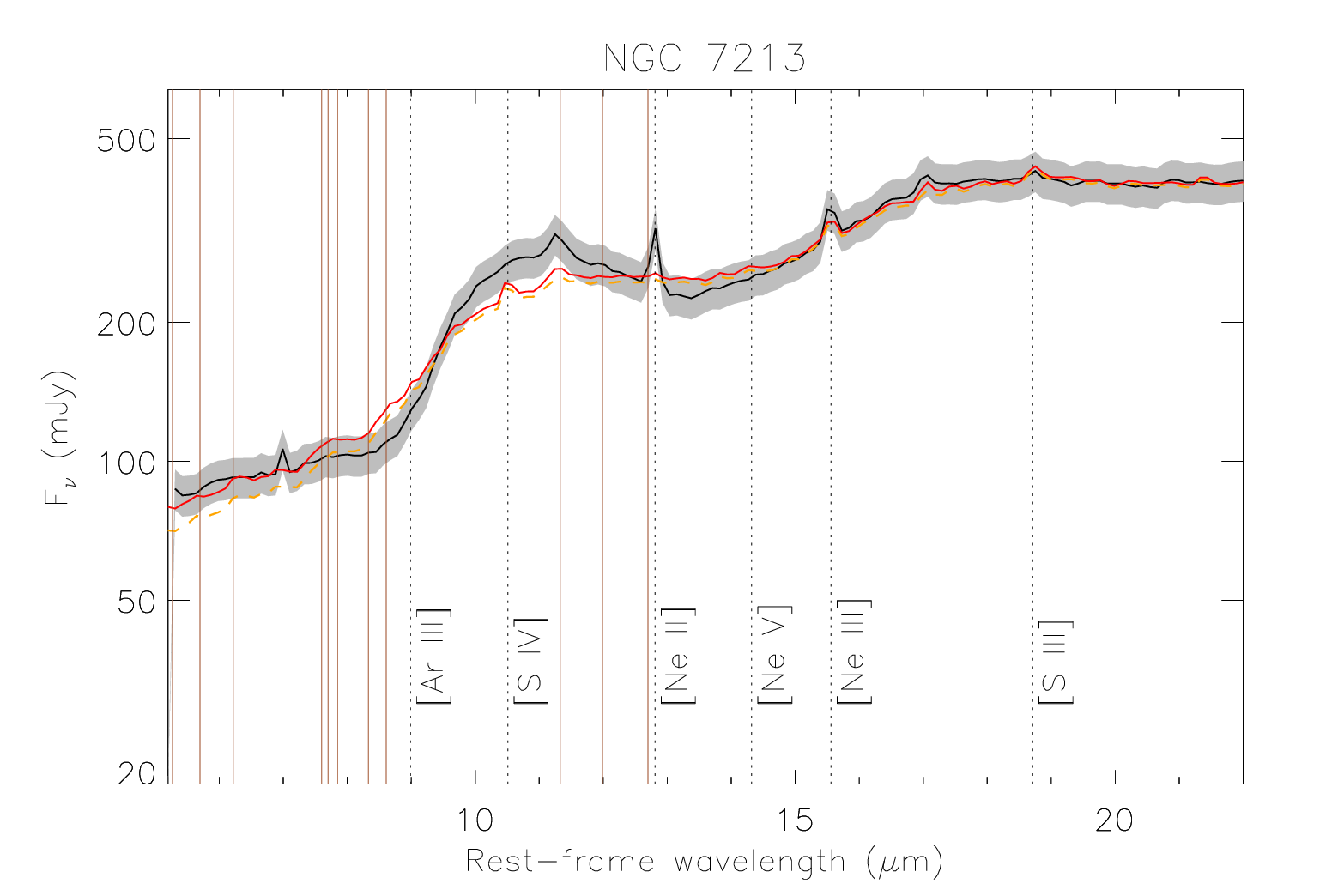}
\includegraphics[width=8.0cm]{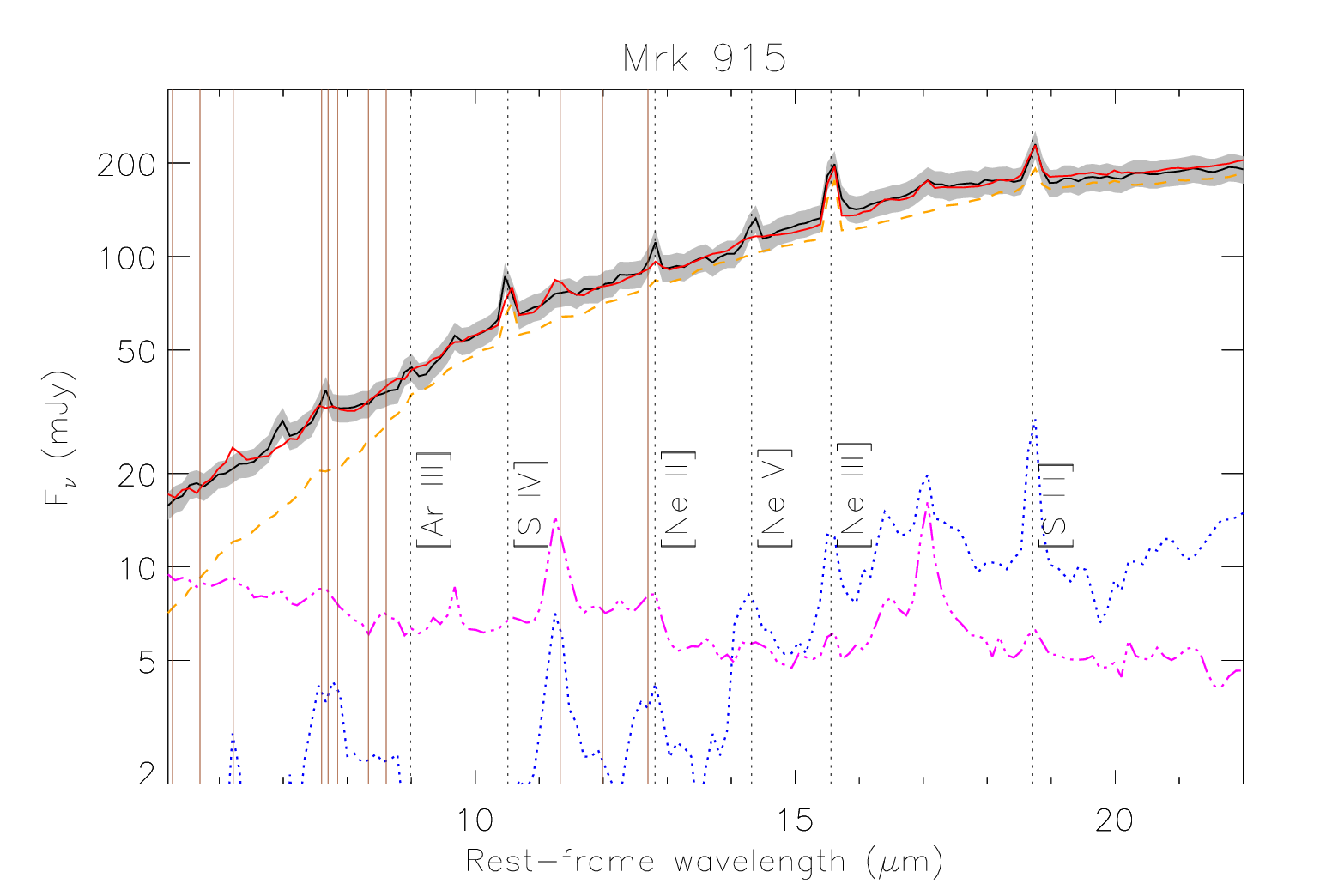}
\par} 
\end{figure*}

\section{Correlation spectra}
\label{B}
\begin{figure*}
\centering
\par{
\includegraphics[width=8.812cm]{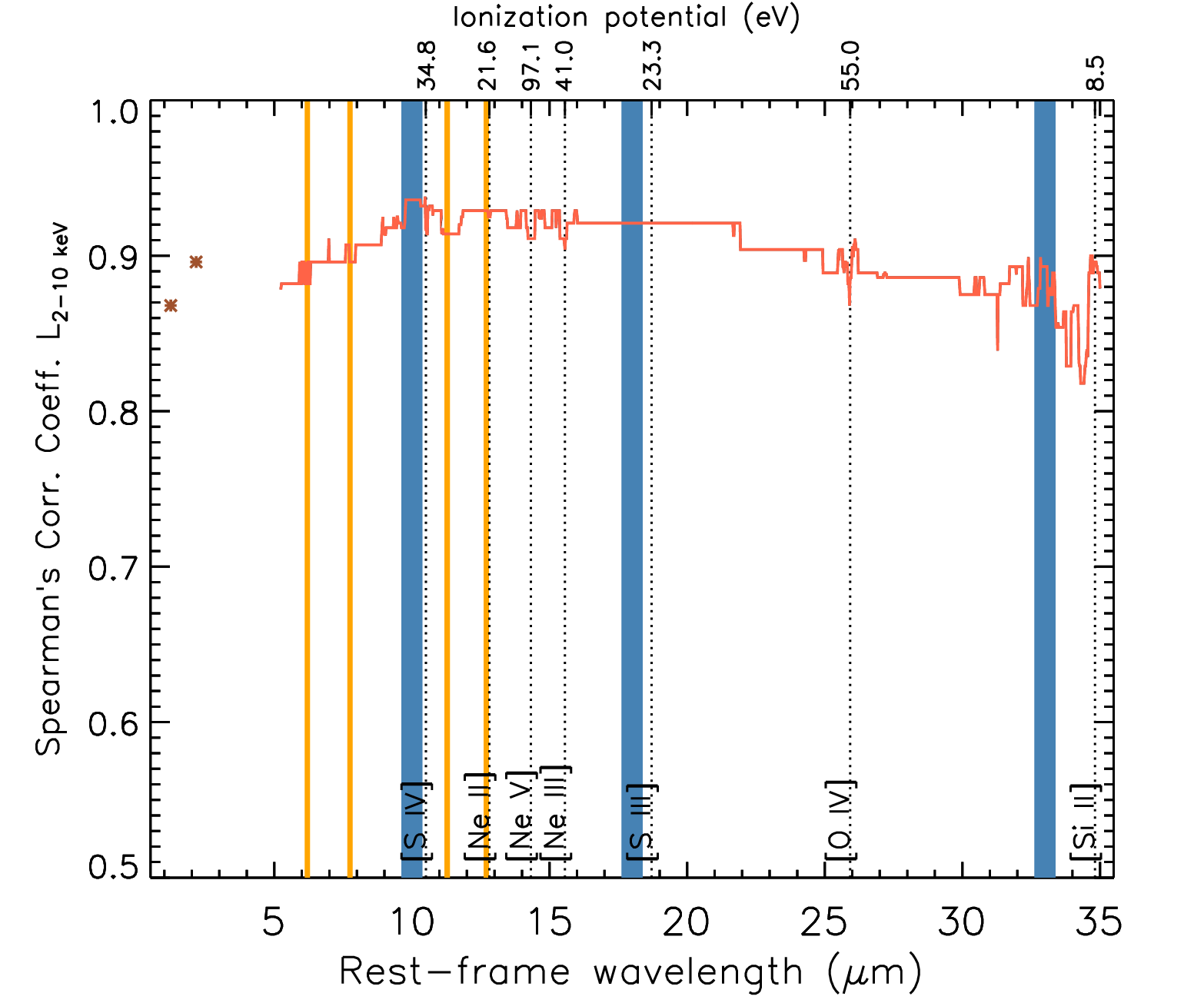}
\includegraphics[width=8.812cm]{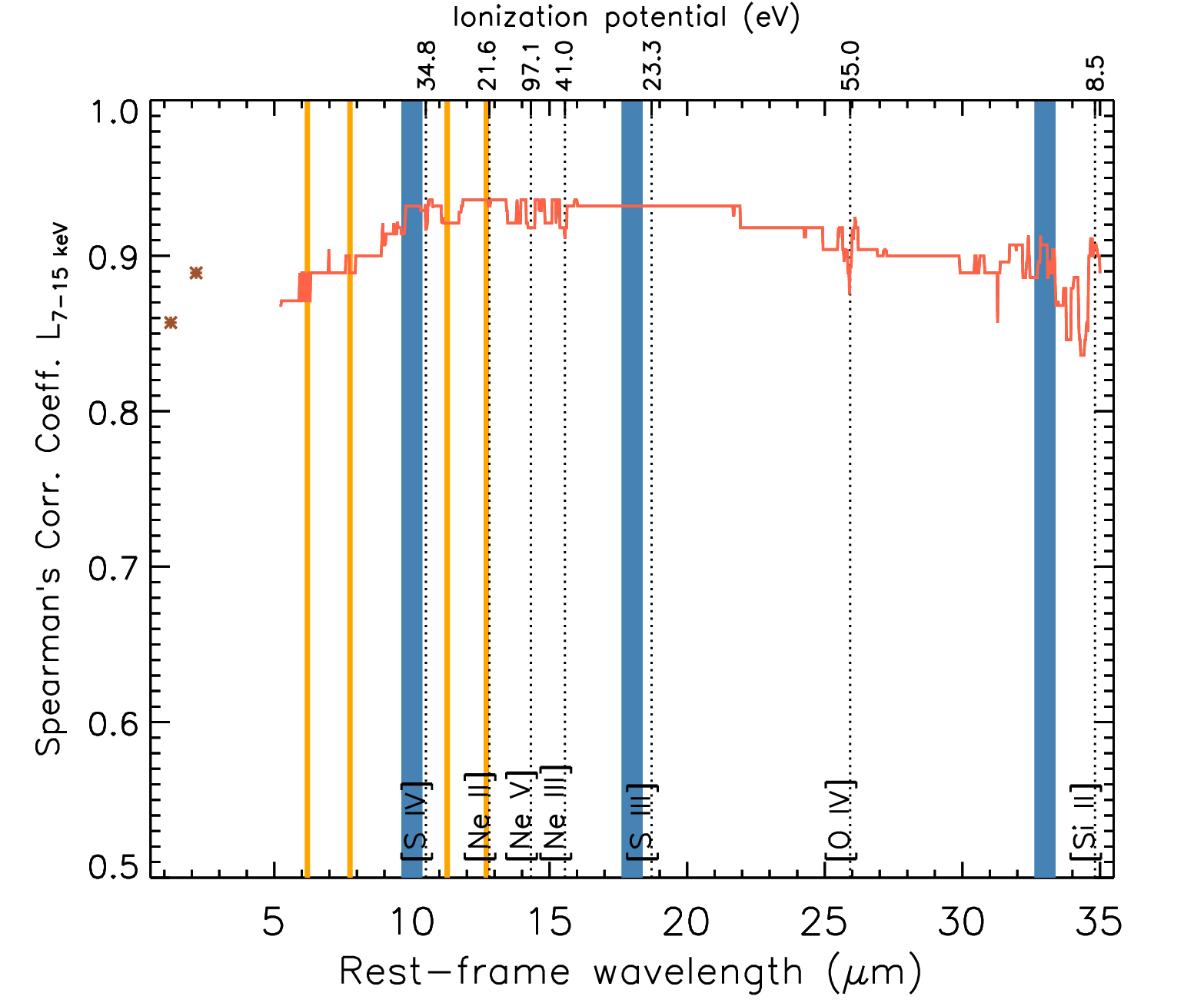}
\includegraphics[width=8.812cm]{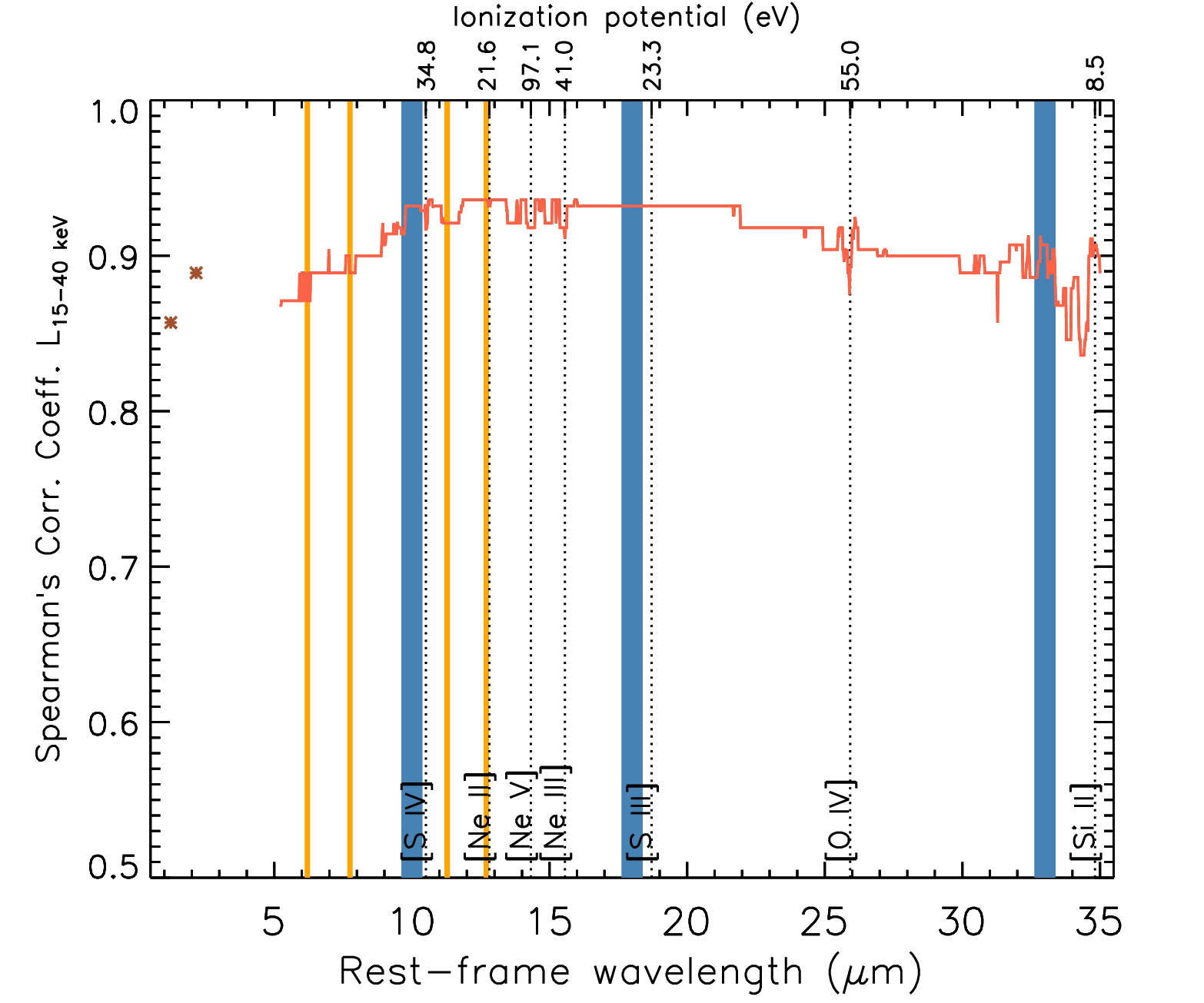}
\includegraphics[width=8.812cm]{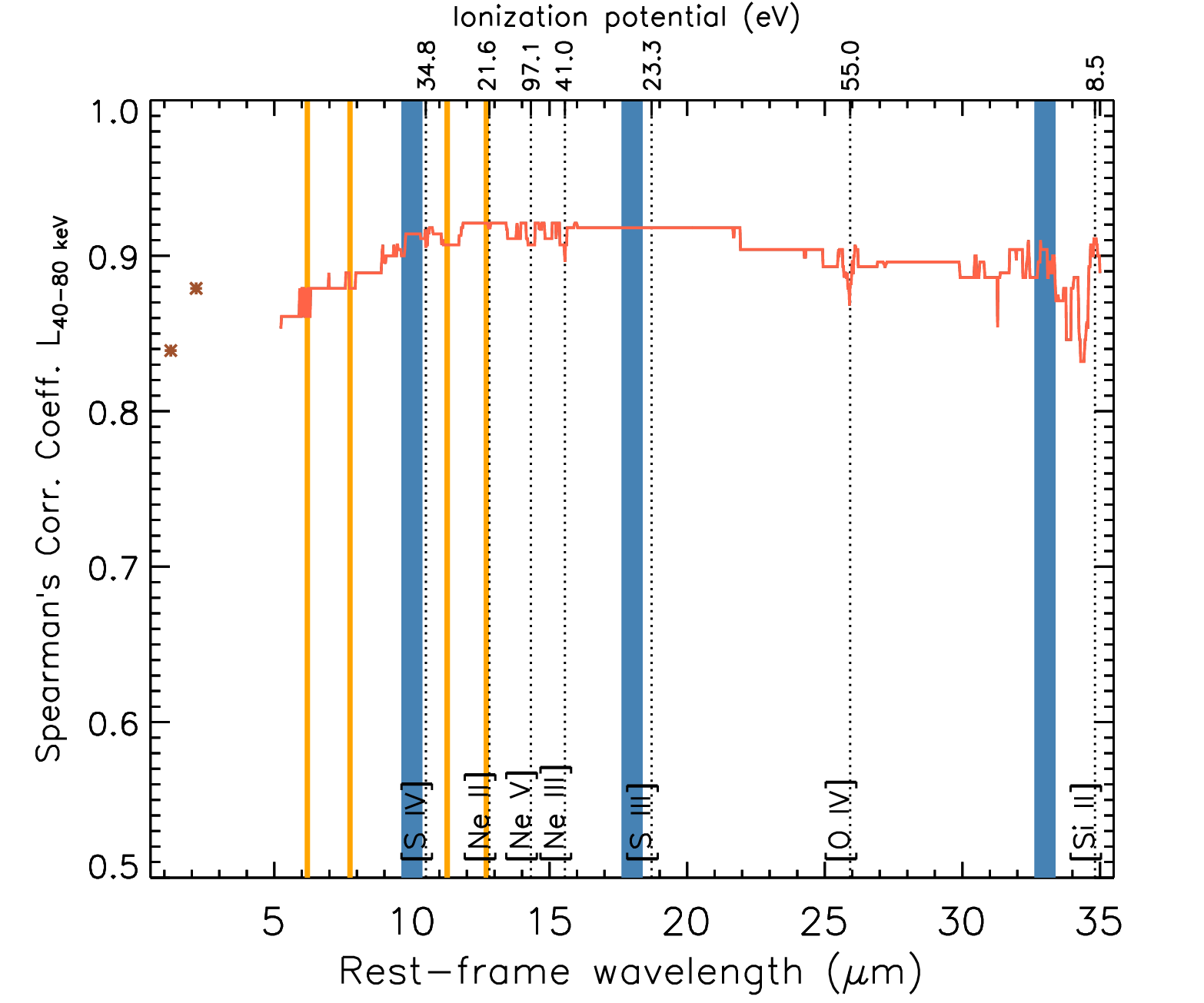}
\par} 
\caption{Same as in Fig. \ref{fig1} but for the subsample of 15 sources without upper limits in the 40-80~keV band.}
\label{figB1}
\end{figure*}

As a sanity check we repeated the correlation spectra for the 2-10, 7-15, 15-40 and 40-80~keV bands, but only using the subsample of 15 sources without upper limits in the 40-80~keV band (see Table \ref{tab2} in Section \ref{xray}). We found that the IR--X-ray correlations are practically identical in all the X-ray bands considered here (see Fig. \ref{figB1}). Therefore, we conclude that the lower significance found for the 40-80~keV IRXCS when the whole sample is considered (see bottom-right panel of Fig. \ref{fig1}) is due to the large number of upper limits included and not a real feature.

\label{lastpage}

\end{document}